\documentclass[proof,hidelinks]{WileyASNA-v1}
\articletype{Article Type}%
\received{00}
\revised{00}
\accepted{00}

\usepackage{placeins}
\usepackage{afterpage}
\newcommand{\mycomment}[1]{}

\begin{document}
\title{Scales of Stability and Turbulence in the Molecular ISM}
\author[1]{Eric Keto}
\authormark{Eric Keto}
\titlemark{Molecular Clouds}
\address[1]{\orgdiv{Institute for Theory and Computation}, \orgname{Harvard University},
\orgaddress{\state{Cambridge, MA}, \country{USA}}}
\corres{Eric Keto}
 \presentaddress{Center for Astrophysics, 60 Garden St, Cambridge, MA 02138}
 
 
 \abstract[Abstract]{ 
We re-analyze the data of the BU-FCRAO $^{13}{\rm CO}$ Galactic Ring Survey (GRS) to understand the dynamics of
the turbulent molecular interstellar medium. We define molecular clouds by their spatial half-power contours of $^{13}{\rm CO}$ integrated intensity, independent of a boundary based on thresholding or tiling. We find properties of hydrostatic equilibrium (HE) and virial equilibrium (VE), the former independent and the latter dependent on time and spatial scales. We suggest that HE is a stationary property of the turbulence and that molecular clouds are high-density regions of a fluctuating component. The gravitational and turbulent kinetic energies within clouds are continuously evolving toward a time-dependent VE with the fluctuating, external, turbulent pressure energy (PE) that can be treated parametrically owing to the shorter time scale for virialization. The average PE is comparable to the pressure of the multiphase ISM at the Galactic mid-plane. Larson's scaling relations analyzed by different statistical methods are not significant. The non-dimensional variances of size, line width, and column density are of comparable magnitude, ruling out the inference of constant column density. Previously unrecognized autocorrelations may have contributed to the apparent validity of the inference.
}


 \maketitle
 \section{Introduction}\label{Introduction}
 
The Boston University-Five College Radio Astronomy (BU-FCRAO) Galactic Ring Survey images
75.4 deg$^2$ of the first Galactic quadrant 
in $^{13}{\rm CO}$ spectra
with 46" angular resolution and Nyquist sampling on a 22" grid \citep{Jackson_2006}.  
Over 6000 molecular clouds\footnote{\citet{Rathborne_2009} define clouds as complexes of clumps within a spatial and velocity range.  We refer to all identifications as clouds. Our definition of a cloud is explained more fully in \S\ref{Definitions} and \S\ref{Methods}.}
 are identified in the survey \citep{Rathborne_2009} with
kinematic distances \citep{Roman-Duval_2009} that enable conversion of the basic cloud properties to 
dimensions of length, mass, and energy \citep{Roman-Duval_2010}.

All the clouds are at least partially resolved by the design of the segmentation algorithm used in \citet{Rathborne_2009} for cloud identification. 
The low observational noise 
\mbox{($\sigma_{\rm T_A} \sim 0.13$ K)} and Nyquist sampling reveal  $^{13}$CO emission extended many times the cataloged radii around the clouds. 
From the $^{13}$CO integrated intensity and the line width, proportional to the column density and the  turbulent velocity dispersion, respectively, we develop
an observational description of the gravitational potential and turbulent kinetic energies
within the clouds and throughout the surrounding molecular interstellar medium (ISM). 
This paper presents the results of our new analysis of the GRS data. 
 
 In agreement with previous results of other surveys \citep{Heyer_2009,MD_2017,Evans_2021}
 we find that 
 the KE and GE energies per unit mass are correlated with an excess of kinetic energy.
We find that $\rm \langle 2KE \rangle / \langle GE \rangle \sim 2.9$, and that
 the turbulent pressure energy per unit mass of the more diffuse molecular ISM around
 each cloud, estimated from the square of the $^{13}$CO line widths, $\rm PE = \sigma^2$, is sufficient to
 complete the virial equilibrium,  $\rm PE \sim 2KE - |GE|$. 
 These results add to the observational evidence for the 
appearance of stability in the molecular ISM. 

Turbulence is pervasive
on all scales in the molecular ISM. 
 Because regions of dynamical equilibrium generally do not exist within fully developed turbulence 
 \citep{VS_1998,Klessen_2005}, 
these observations of stability seem at odds with
theories of molecular clouds as 
turbulent eddies or compressed regions between colliding flows   
\citep{Elmegreen_1993,MLK_2004}. 
 Why the clouds 
 appear to be in virial equilibrium  
 has not been explained.
 
 This paper suggests that the resolution of the paradox of virial equilibrium within turbulence is found by consideration of the
 different time scales for the evolution of the three terms for energy KE, GE, and PE.
 The dynamical time scale
 for the turbulence at any scale is approximately its crossing time or eddy turnover time, 
 $T_X = R/\sigma$ \citep{Kolmogorov_1941,Gammie_1996}. 
 If the 
 turbulent velocity dispersion, $\sigma$, scales with a power of the length scale, 
$R$, less than one \citep{Heyer_2009,Padoan_1995},
then the turbulence on smaller spatial scales evolves with a shorter time scale than on longer spatial scales.
 
 The rate of virialization within a cloud is then more rapid than the rate of change
 of the external PE due to turbulent fluctuations around the cloud.
 In the snapshot in time
 represented by an observation,  the clouds are in a time-dependent, continuously 
 evolving virial equilibrium
 within a more slowly evolving external pressure that can be considered parametrically 
 in the virial theorem applied to each cloud.

In section \S\ref{Larson}, we consider 
the two empirical scaling relations between cloud size, line width, and column density proposed in \citet{Larson_1981},
possibly indicative of properties of the turbulent, molecular ISM \citep{Kritsuk_2013}.
Our results generally confirm the previous GRS study \citet{Heyer_2009}.
Additionally, we evaluate the statistical significance by comparing the mean square errors of the regressions
of the cloud properties
with their variances and find that neither correlation is significant.
A comparison of the variances in non-dimensional units shows that the independence of
column density and cloud size cannot be interpreted as implying constant column density.
Rather the two quantities have uncorrelated log normal distributions.

In \S\ref{Complexity}, we define a complexity parameter to measure the departure from the circular shape expected for
a cloud in the spherical minimum-energy configuration of hydrostatic equilibrium. We find 90\% of the clouds have
shapes inconsistent with hydrostatic equilibrium.

In \S\ref{PDF},  we compare the column density probability distribution functions (PDF) inside and outside the clouds. 
Both are close to log normal.

The Discussion \S\ref{Discussion} suggests that the conditions in the molecular ISM, as described by the GRS,
are suitable for the growth and fragmentation of molecular clouds by a turbulent cooling instability \citep{Keto_2020}.
       
 \section{Definitions}\label{Definitions}
 
 We consider molecular clouds and the virial theorem differently than many previous observational studies. 
 This section explains the definitions of molecular clouds, cloud size, and time-dependent equilibrium. 
 
 \subsection{Definition of a molecular cloud}\label{Cloud_def}
 
  We define a cloud as a region of higher column density identified
  by a peak in the $^{13}$CO integrated intensity that is at least twice the level 
 of the azimuthally-averaged  intensity around the peak. 
 This defines a cloud outward from its center without reference to a boundary 
 specified as an intensity threshold or as a tiling from a segmentation algorithm.
 Higher-density regions within larger clouds (sometimes called clumps) are identifiable
 with the same definition applied to any local peaks and their surrounding intensities.
 We refer to all over-densities meeting our definition simply
 as  clouds.
 
 \subsection{Definition of the cloud size or radius}\label{Cloud_radius}
   
We define
 a characteristic length (radius) for each cloud as twice the HWHM of the radial intensity profile azimuthally
 averaged around the peak.
  The motivation for the
 factor of two follows from the power-law density structure of an idealized cloud described by the 
 Lane-Emden equation. 
 Similar to the description of an exponential curve,
 we identify an outer region in the density profile that
 may be approximated as linear with respect to an inner non-linear region where the density profile rises steeply.
 The break point occurs at approximately 2 HWHM.  The properties of the molecular ISM are continuous
 across the defined radius (\S\ref{profiles}).
  
 We use the length scale of the cloud to separate the length scales of the  
 turbulence into the smaller scales inside
 the cloud and the larger scales outside. Results of the analyses that depend on this separation,
 for example virial equilibrium in an external pressure, are not sensitive to the exact definition of the length scale
 owing to the scale-free nature of turbulence and the 
 continuity of properties within the molecular ISM (\S\ref{Hydrostatic}).
    
 \subsection{Definition of time-dependent virial equilibrium}\label{def_VE}

The time-dependent virial theorem in units of energy per unit mass including an external pressure 
 \citep{Field_2011,Spitzer_1978} is,
 \begin{equation}
\delta(t) = 3M\sigma^2(t) 
- \beta \frac{G M^2}{R(t)}
- 4\pi R(t)^3 P_{\rm ext}(t),
\end{equation}
where 
 $\delta(t) $ is a measure of disequilibrium at time $t$,
 $\sigma^2(t)$ is the average turbulent velocity dispersion,
 $P_\mathrm{ext}(t)$ is the external pressure ,
 and $\beta$
 depends on the density distribution within the cloud,
 3/5 for uniform or 0.732 for hydrostatic used here.
 With a Lagrangian
 interpretation,
the mass $M$ is constant within a time variable radius $R(t)$.

This can be rewritten per unit mass in terms of observable quantities as,
 \begin{equation}\label{ve1}
 \delta(t) =  \sigma(t)^2 - \Gamma\Sigma(t) R(t) - \frac { \pi \Sigma( R(t) ) } { 4 \Sigma(t) } \sigma^2(R(t),t)
 \end{equation}
 where 
 $\Sigma( R(t) )$ and  $\sigma( R(t) )$ are the column density and 
 the velocity dispersion outside the virial boundary $R(t)$. 
 The constant $\Gamma = \pi\beta G/3$. The
 factor $\pi/ 4$ in the last term assumes that the virial boundary is inside the power-law envelope
 of the self-gravitating cloud
 where the density profile
 scales as $\rho \propto r^{-2}$. In this case, the integral for the projected
 density may be solved analytically,  and the line-of-sight average density
 outside the cloud is $\rho( R(t) ) = \Sigma( R(t) ) /( \pi R(t)) $. 
 The average density inside
 $R(t)$ is
 $\langle \rho \rangle = \Sigma / (4 R(t))$.
    
Because we lack observational data on the magnetic field energies in the GRS clouds, 
a separate term for the magnetic energy is not included in equation \ref{ve1}. To the level of accuracy of our observational analysis,
one could assume that the turbulent magnetic energy is in equipartition with the observed turbulent kinetic energy. 
In this case, the external pressure energy derived from the GRS data (\S\ref{Properties})
would be higher by less than a factor of 2. 

 \section{Methods}\label{Methods}
 
 This section on methods describes the procedures to  select the subset of clouds suitable for analysis, to co-add radial profiles
 across different length scales, and to calculate the cloud properties such as
 column density, line width, and external pressure.  Readers more interested in the results may proceed to \S\ref{Results} and refer back here as 
 necessary. 
 
 \subsection{Identification of the Clouds}\label{Cloud_ID}
 
 We rely on the positions and velocities in the catalogs of \citet{Rathborne_2009} for
 preliminary identification of individual clouds. 
 Not all the clouds are suitable for our analysis. 
 
   We define the following filters:\\
 \noindent (1) No distance. Not all the clouds in the catalog of \citet{Rathborne_2009} 
 have an estimated distance in the catalog of \citet{Roman-Duval_2009}.\\
 (2) Overlapping spectral lines. Clouds with spectral lines that overlap within the spectral 
 HWHM of the VLSR are rejected. Individual spectral lines are identified by a peak-finding algorithm that traverses the spectrum and 
 identifies a separate spectral line as a local peak that is five times the local noise level above the last local minimum. \\
(3) Duplicate identification.  Two clouds that have peak integrated intensities within 3/4 of the survey angular resolution
and VLSRs within each other's line widths are the same cloud.\\
(4) Peak integrated intensity below 1 K km s$^{-1} $pixel$^{-1}$ (S/N <  $\sim 7$). 
While a peak may be identifiable, the intensities off-peak are too low for our analysis. \\
(5) No half-power width. The peak intensity might not be above twice the level of the surroundings; or,
the radial profile might be too complex. \\
(6) Confusion with another cloud. If a local intensity peak is located within another cloud whose emission is steeply rising in one direction, the peak
may have a half-power level, but the majority of pixels within the HWHM may have intensities above the peak or a single pixel may have
an intensity more than 1.5 times the peak. \\

\vbox{
\noindent (1) Number of clouds missing a distance: 357 \\
(2) Number of clouds with overlapping spectral lines: 257 \\
(3) Number of clouds with duplicate identifications: 878 \\
(4) Number of clouds with peak integrated intensity < 1 K: 229 \\
(5) Number of clouds with no half-power point: 402 \\ 
(6) Number of clouds with confusion: 170 \\
Number of clouds selected for analysis: 4190 out of 6155. Some clouds are rejected for more than one reason.
}

\subsection{Properties of Individual Clouds}\label{section_cloud_ID}
Previous studies of the GRS data \citep{Heyer_2009,Roman-Duval_2010} derived the molecular gas
column densities from $^{13}$CO and $^{12}$CO excitation temperatures
to minimize optical depth effects. 
They used the $^{12}$CO data from the earlier University of Massachusetts-Stony Brook Galactic Plane Survey (UMSB) \citep{Solomon_1987}.
We find that we can reliably match the column densities in \citep{Heyer_2009} with the $^{13}\rm CO$ conversion factor of 
22 $\rm M_\odot pc^{-2} / K\ km\ s^{-1} pc^2$, suggested in \citet{Phiri_2021}.
We compute the spectral line width
from the second moment of the $^{13}$CO spectral line. 

We calculate the column density and line width for each pixel in a $128 \times 128$ pixel box ($96\times 96$ arc minutes) 
around the peak integrated intensity of each cloud. Pixels with intensity less than five times the observational noise
 are ignored. 
The column density and velocity dispersion of a cloud are taken to be the averages, 
respectively, of the quantities per pixel within a circular radius
of 2 HWM (\S\ref{Cloud_radius}) around the peak. 
The mass is the sum of the column densities per pixel. 

\subsection{Co-added radial profiles}\label{Co-addition}

The radial profiles of the $^{13}$CO integrated intensity and line width of individual clouds can be 
co-added by scaling the lengths of both profiles by the HWHM of the column density
profile and normalizing their respective line intensities or line widths.  
The variation of the line width per pixel is noisier
than that of the intensity, so for the purpose of normalization, we define the line width at the cloud center as the average line width within half
the HWHM.

There is often emission from more than one cloud within our 128 x 128 pixel analysis box around each cloud and within the spectral window 
(VLSR$_{\rm cloud} \pm$ FWHM). 
This emission is included in the azimuthal averaging of the radial profiles for the individual clouds (\S\ref{Properties}).  
Because azimuthal
 averaging is inherent in our definition of a cloud, the effect of a nearby cloud on the gradients is diluted by the fraction of its 
 angular intercept and by the separation squared. 
For the calculation of the sample-average radial profiles, we exclude clouds whose radial profiles are strongly affected by
other clouds. We exclude 507 clouds with profiles where the normalized intensity
exceeds 0.4 beyond 3 HWHM.  The co-added average profiles include 3683 clouds. 

 The observed azimuthally-averaged radial profiles around the individual clouds extend to the boundaries of our analysis box
 around each cloud, 48 arc minutes, 
regardless of the size of the cloud. 
Clouds with larger HWHM measured in arc minutes do not extend as far in non-dimensional units as those around smaller clouds. 
Thus the averaging at large non-dimensional radii includes fewer clouds and the average profile becomes noisier with distance.
After rescaling and co-adding we have good data out to a non-dimensional radius of 20 HWHM.


\begin{figure}[hbt!]
\begin{tabular} {c}
\includegraphics[width=3.65in,trim={0 0.4in 0.5in 0.4in},clip]{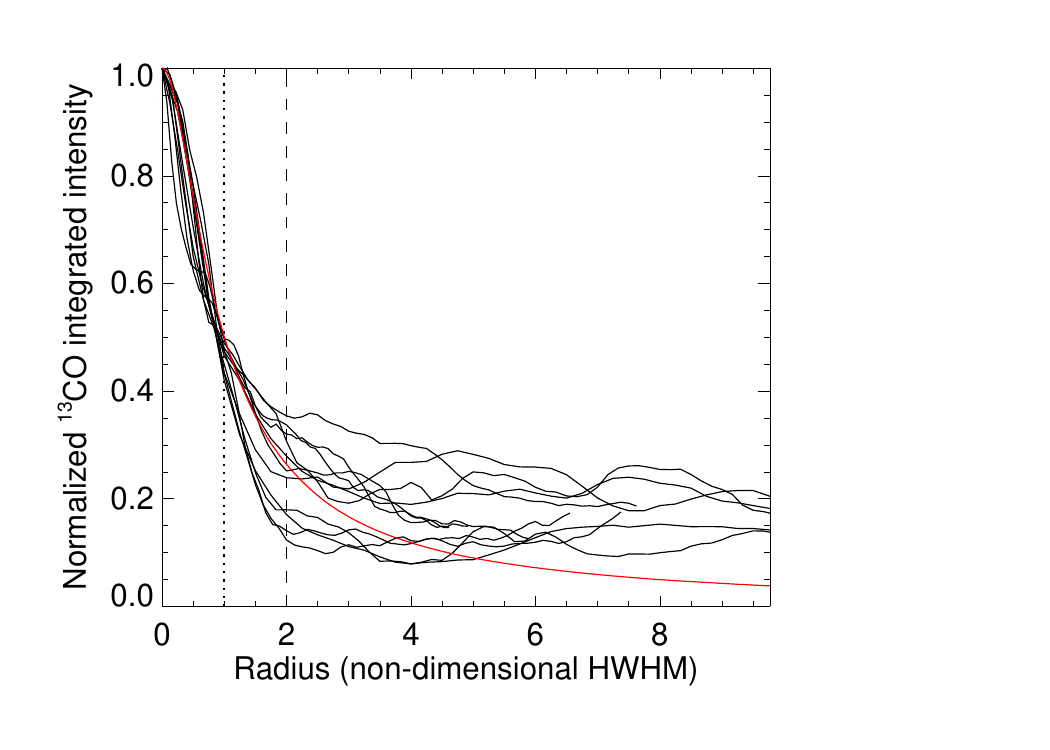} \\
\includegraphics[width=3.65in,trim={0 0.4in 0.5in 0.4in},clip]{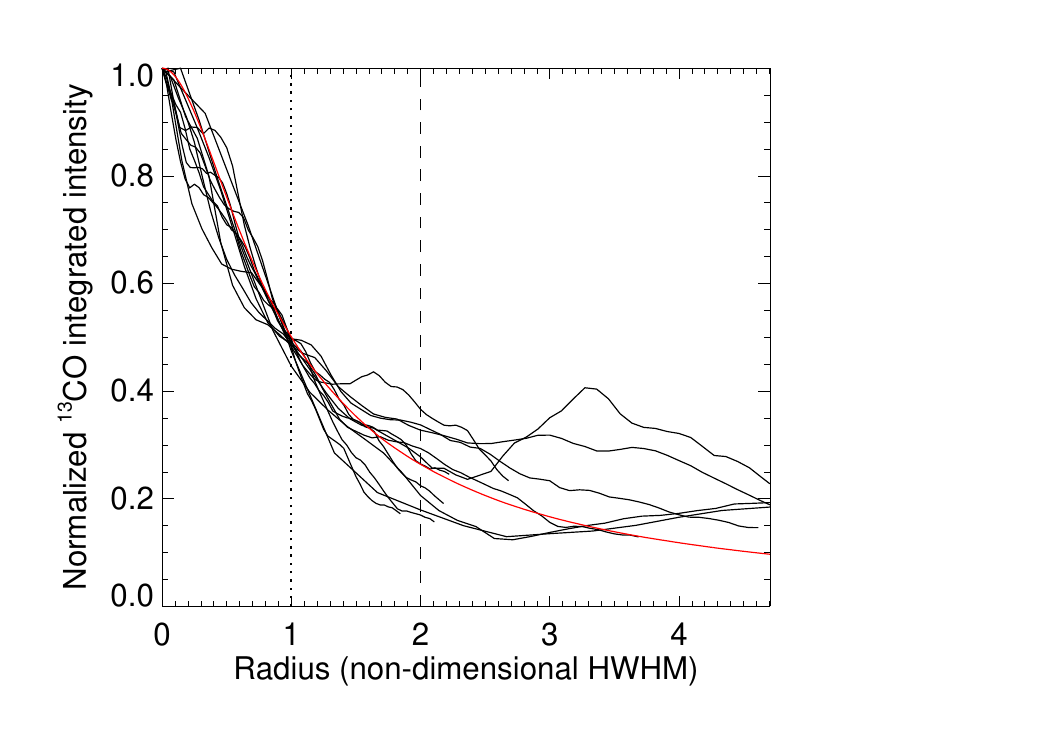} \\
\includegraphics[width=3.65in,trim={0 0.4in 0.5in 0.4in},clip]{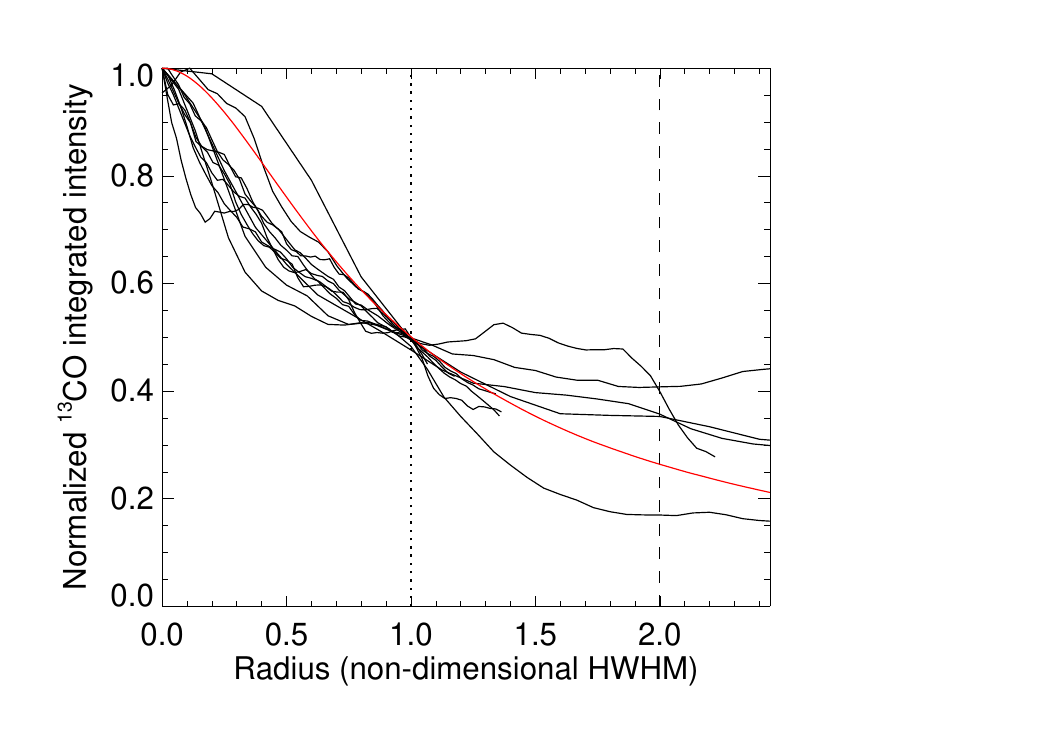} \\
\end{tabular}
\caption{
Thirty individual azimuthally-averaged radial profiles of $^{13}$CO integrated intensity (K km s$^{-1}$ pixel$^{-1} $).
{\it Top} to {\it bottom} show ten profiles each for clouds that have small, medium, and large sizes
with respect to each other.
The vertical, dotted and long-dashed lines indicate the HWHM and the 
length scale of the cloud defined as 2 HWHM. The red line shows the column density of 
hydrostatic equilibrium (\S\ref{Hydrostatic}), same as in figure \ref{coadded_profiles}\!\! ({\it right}, blue, long-dashed line).
}
\label{individual_profiles}
\end{figure}

\begin{figure}[hbt!]
\begin{tabular} {c}
\includegraphics[width=2.90in,trim={0 0.4in 2.2in 0.4in},clip]{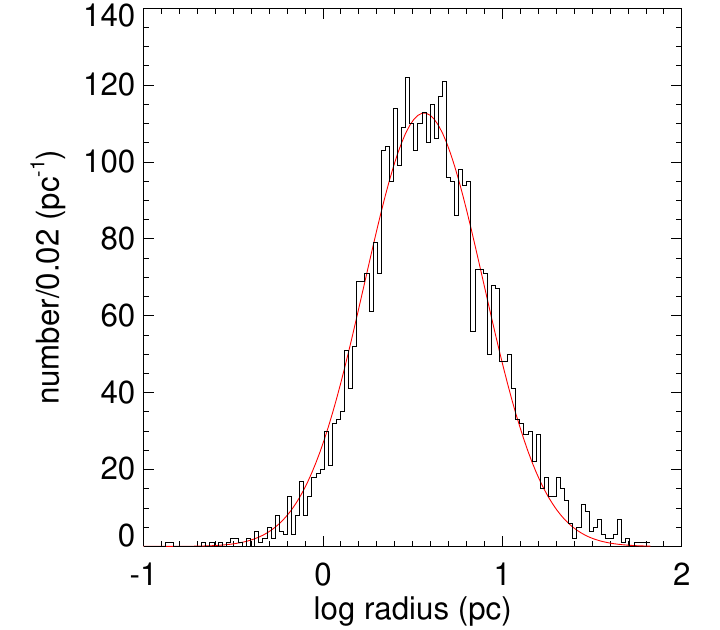} \\
\includegraphics[width=2.90in,trim={0 0.4in 2.2in 0.4in},clip]{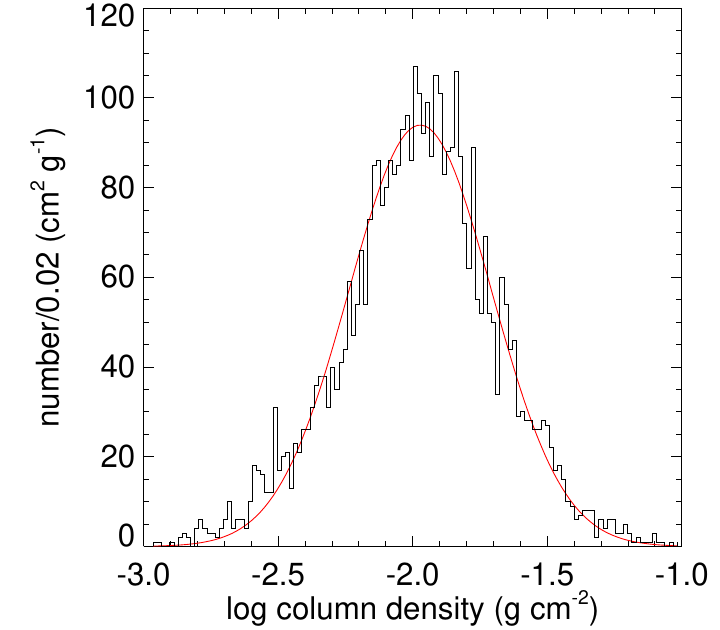} \\
\includegraphics[width=2.90in,trim={0 0.4in 2.2in 0.4in},clip]{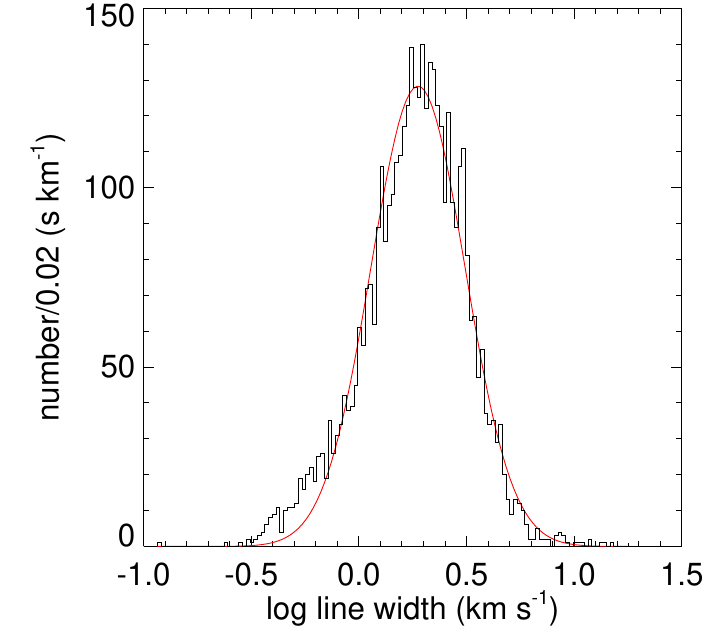} \\

\end{tabular}
\caption{
Histograms showing the log normal distributions of radius, column density, line width.
Means and standard deviations are listed in
table \ref{table_lognormals}.
}
\label{histograms}
\end{figure}

\begin{figure*}[!ht]
\begin{tabular}{ll}
\includegraphics[width=2.8in,trim={0.3in 0.2in 1.2in 0},clip] {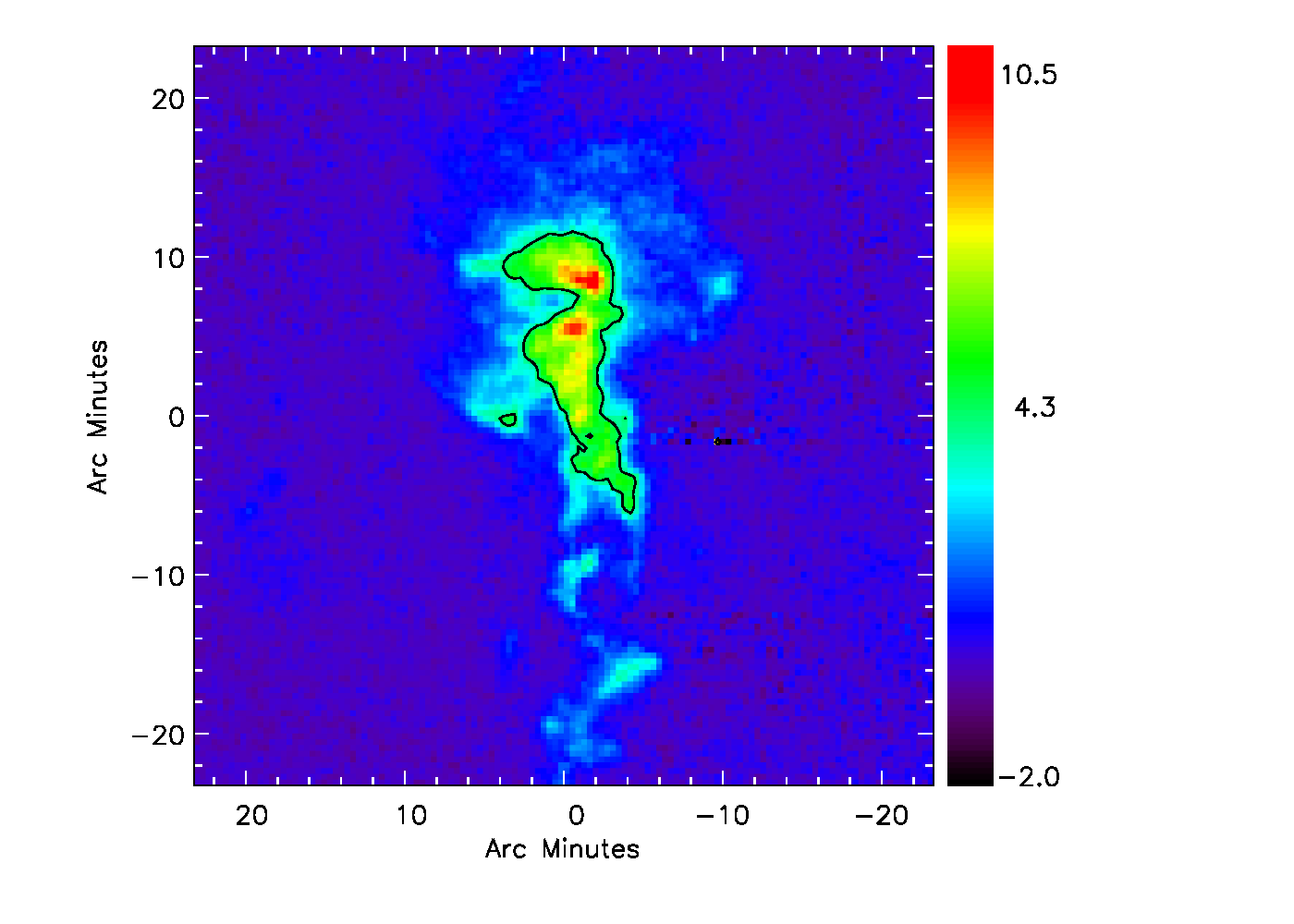} & 
\includegraphics[width=3.0in,trim={0.3in 0.2in 1.2in 0.3in},clip]{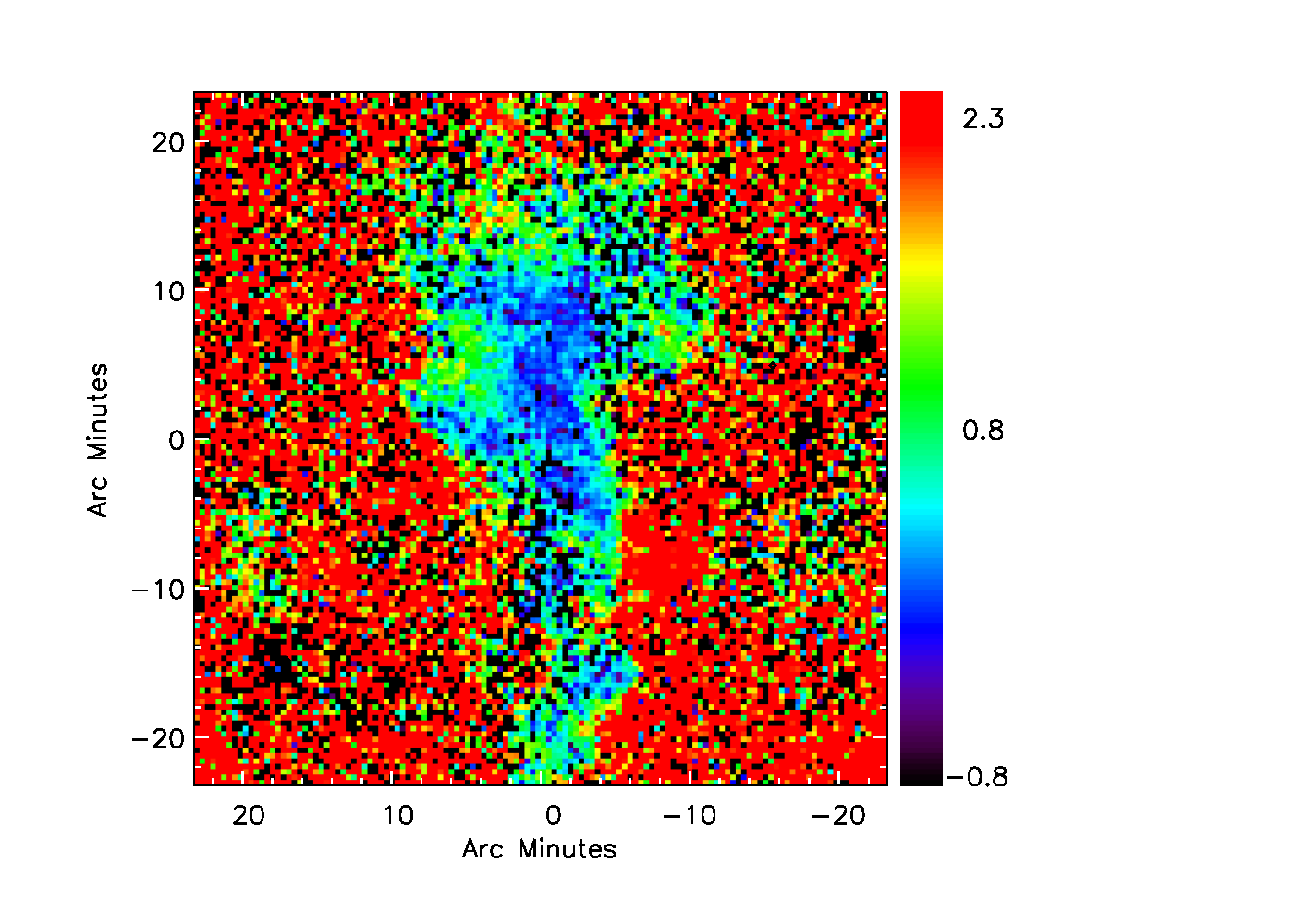}\\ 
\includegraphics[width=3.32in,trim={0.3in 0.2in 0.0in 0.3in},clip] {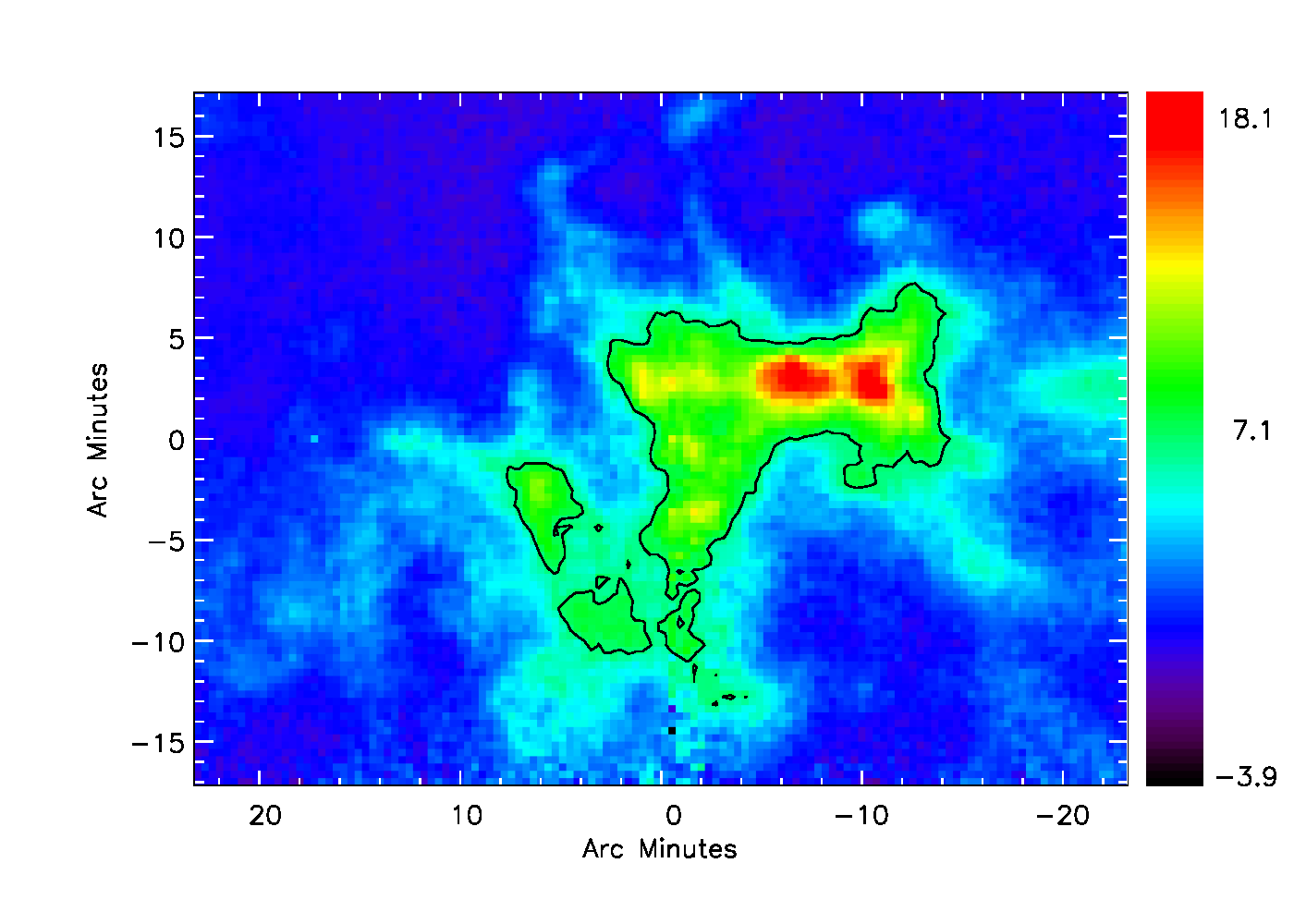} & 
\includegraphics[width=3.32in,trim={0.3in 0.2in 0.0in 0.3in},clip]{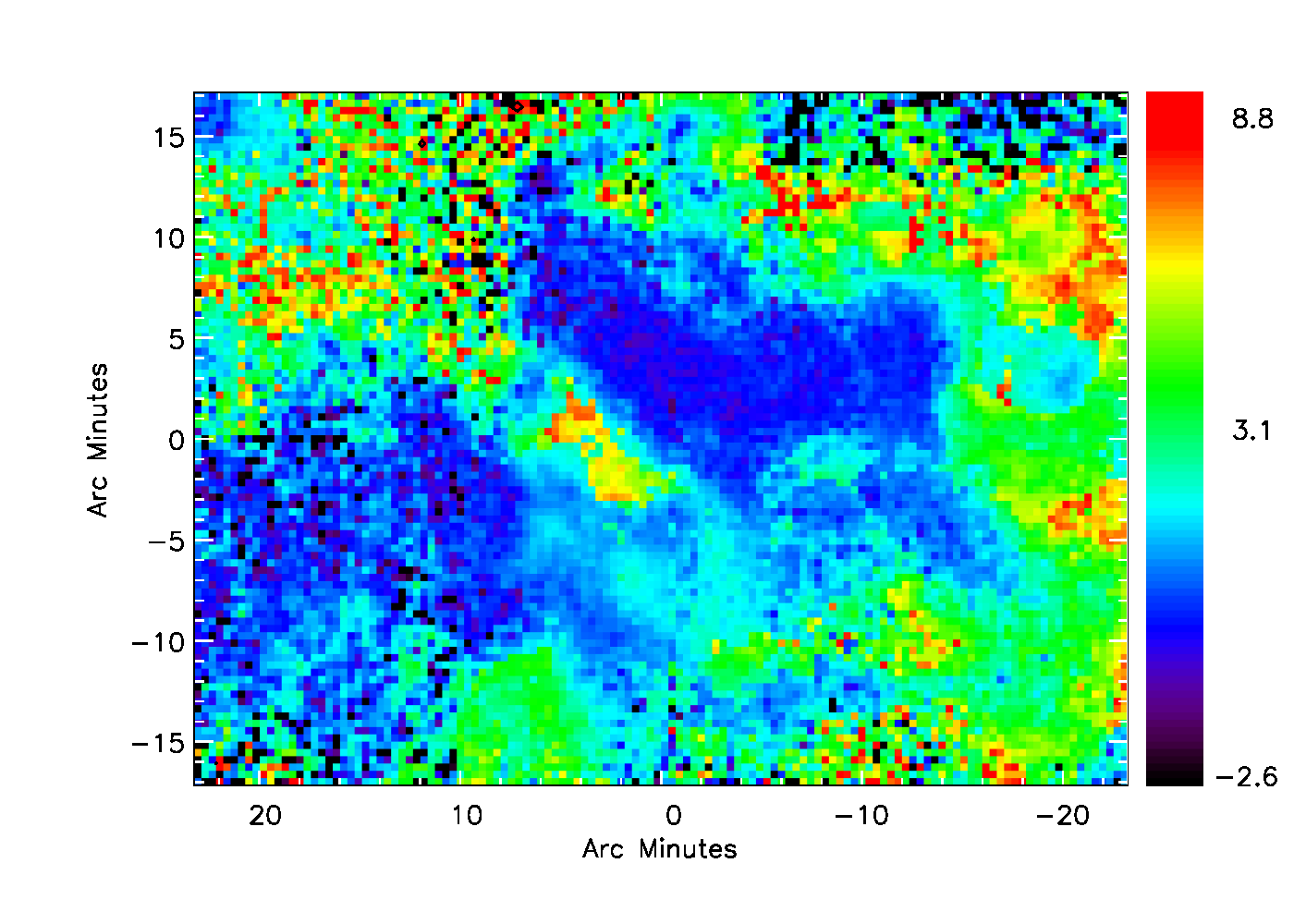}\\ 
\includegraphics[width=2.8in,trim={0.3in 0.2in 1.2in 0},clip] {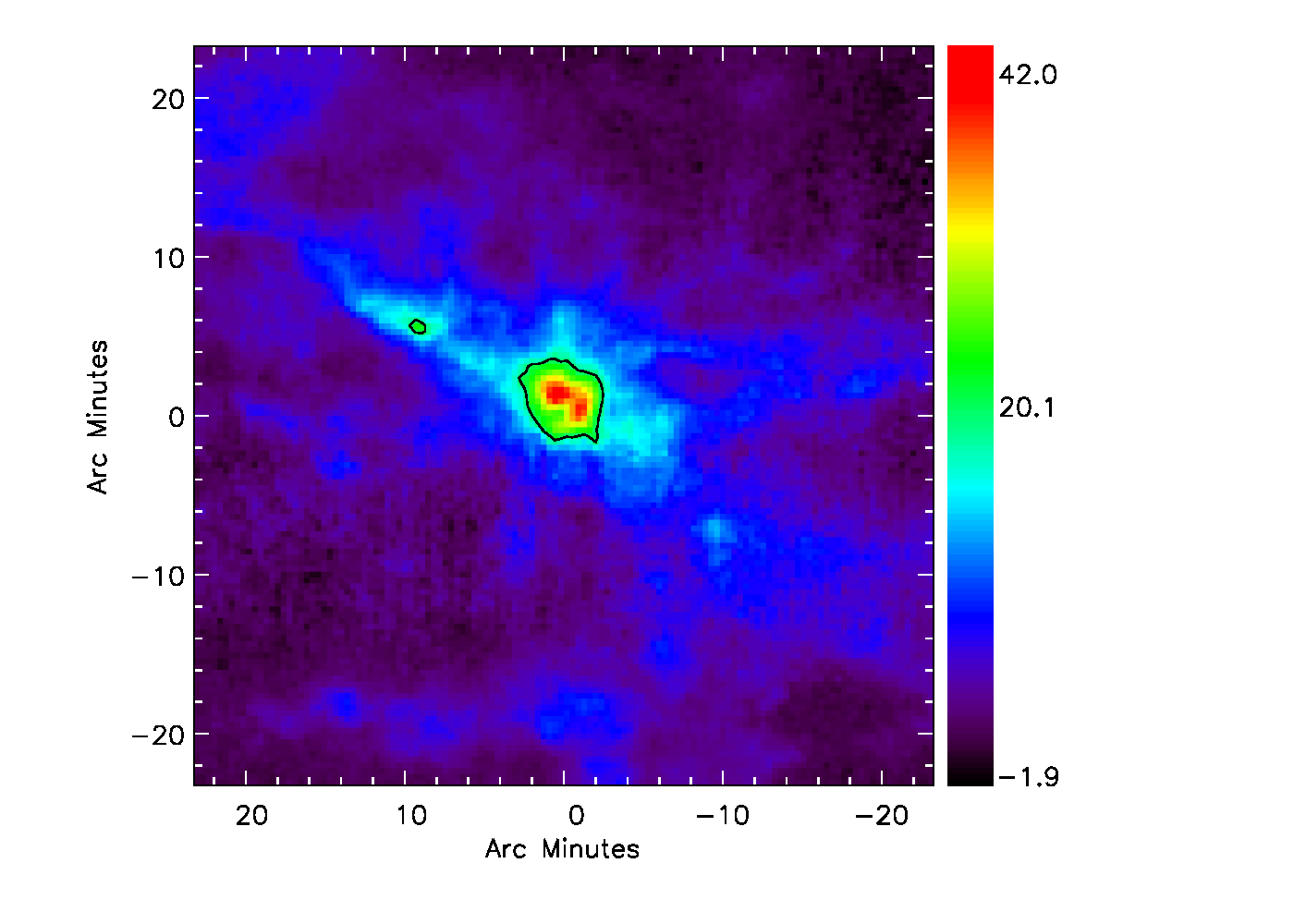} & 
\includegraphics[width=2.9in,trim={0.35in 0.2in 1.35in 0.3in},clip]{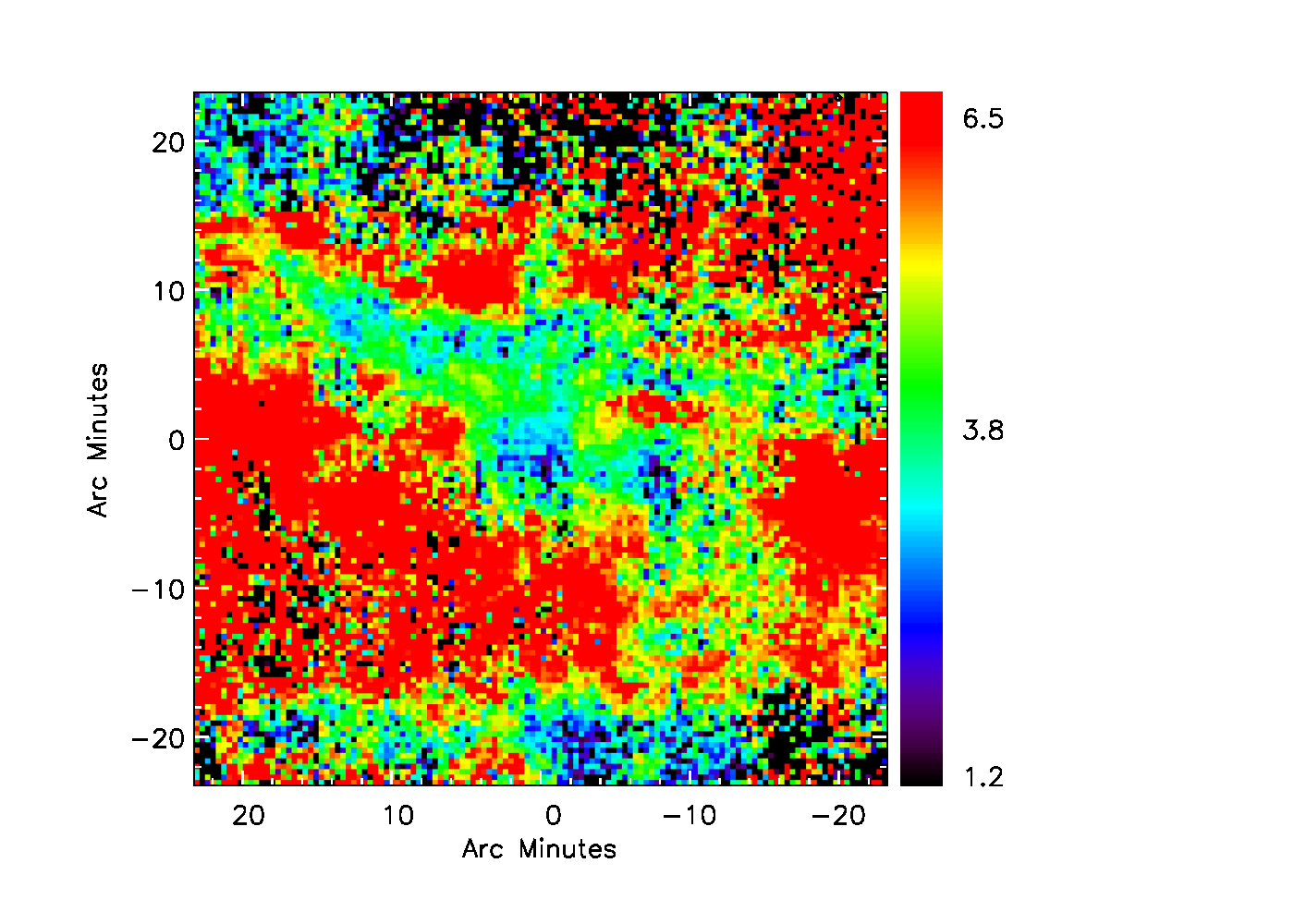}\\ 
\end{tabular}
\caption{
({\it Left}) The $^{13}$CO integrated intensity (K km s$^{-1}$ pixel$^{-1}$) and ({\it right}) line width (km s$^{-1}$) with
a color bar indicating the values in these units, respectively. 
From {\it top} to {\it bottom}, the clouds shown are clumps 54, 64, 70 from the clumps catalog of \citet{Rathborne_2009}. The 
single contour ({\it left}) is drawn at the half-power level of the peak integrated intensity within the contour. The cloud length scale is
twice the circularly averaged HWHM. 
}
\label{paired_maps}
\end{figure*}

\section{Results}\label{Results}

\subsection{Radial Profiles of Individual Clouds}\label{profiles}

Figure \ref{individual_profiles}\!\! shows 
radial profiles of column density for 30 example clouds in three groups of ten 
for smaller, medium-sized, and larger clouds with respect to each other.
The profiles are normalized to units of the HWHM (\S\ref{Co-addition}). 
Comparison shows that for clouds of all sizes,
our arbitrary definition of the cloud length scale at 2 HWHM (\S\ref{Cloud_radius})
is a reasonable
division between the steeply rising column densities within the clouds and the flatter surrounding 
column densities  
although the profiles are continuous. 
The profiles appear to indicate approximate hydrostatic equilibrium (\S\ref{Hydrostatic}) with 
random perturbations that can be ascribed to turbulence.  

\subsection{Properties of the Individual Clouds}\label{Properties}

Figure \ref{histograms}\!\! shows histograms of the three variables, the radius, Gaussian line width, and
column density and their best-fit log normal functions with means and standard deviations listed in
table \hbox{\ref{table_lognormals}\!\!.}
The median values of our cloud properties are listed in table \mbox{\ref{table_medians}\!} along with those
calculated from the catalogs of \citet{Rathborne_2009} following the same procedure we use to convert observational to
physical units but with their clouds defined by their segmentation.

The median radii of our clouds are larger than those in \citet{Rathborne_2009},
while the median column densities and velocity dispersions are the same.
The segmentation algorithm used in \citet{Rathborne_2009}
 tiles the emission into
 clouds that intersect only on their boundaries.  A cloud with multiple intensity peaks may be tiled into
 smaller clouds around each peak.  Our clouds, defined to study the continuous gradients of
 column density and turbulent energy, would, by comparison, extend  
 across these truncations imposed by the tiling
 to result in overlapping clouds that share pixels.

 \subsection{An average hydrostatic equilibrium across the scales of turbulence}\label{Hydrostatic}

Most of the clouds have $\rm ^{13}CO$ line widths per pixel 
that are lower in the cloud centers and lower than in the surrounding gas. 
This can be seen in two ways: in maps of the spectral line width and in the sample-average
radial profile of the line width. 

Figure \ref{paired_maps} shows pairs of maps for several clouds. The map on the left is the $^{13}$CO integrated intensity 
(K km s$^{-1}$ pixel$^{-1}$)  that is proportional to the column density after accounting for distance. 
The map on the right shows the spectral line width (km s$^{-1}$). 
The pairs show a negative correlation between the column density and the line width.

\begin{figure*}[!ht]
\begin{tabular} {p{3.5in}c}
\includegraphics[width=3.5in,trim={0.3in 0 2.1in 0},clip] {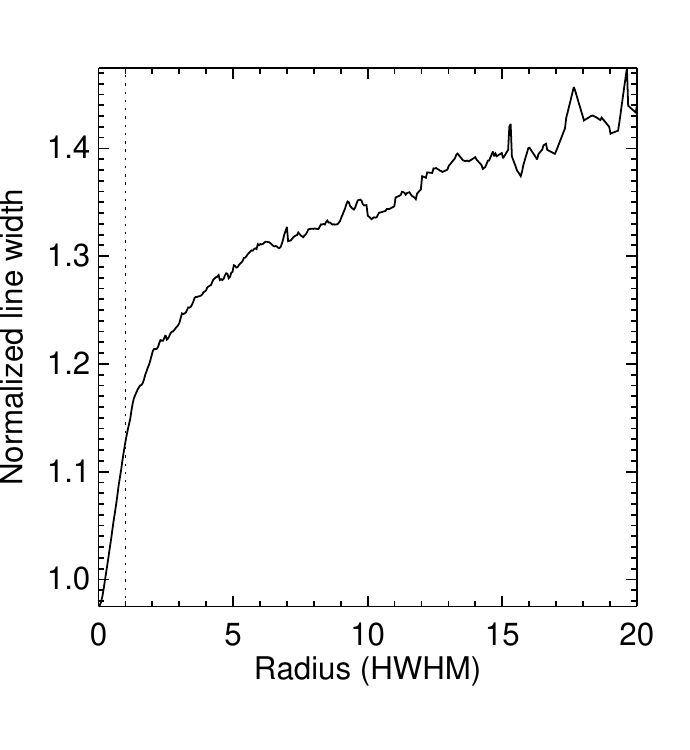} &
\includegraphics[width=3.5in,trim={0.3in 0 2.1in 0},clip] {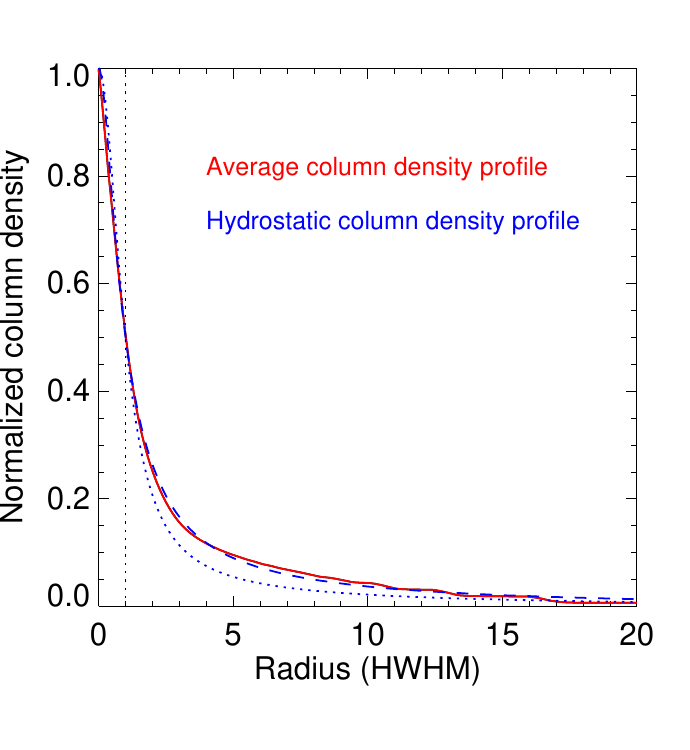}
\end{tabular}
\caption{
{\it Left:} Average non-dimensional line width profile of 3683 clouds (\S\ref{Co-addition}). 
{\it Right:} Average non-dimensional column density profile. 
In both
figures, the vertical dashed line indicates $\rm HWHM = 1$. 
The observed profile of column density is
shown {\it right} as a red line. Two profiles of the column density in hydrostatic equilibrium are also shown
({\it right})
in blue with the dotted line for a non-dimensional isothermal temperature and the long-dashed line for 
a non-dimensional temperature proportional to the square of the line width ({\it left}).
}
\label{coadded_profiles}
\end{figure*}

Not all the individual clouds
in the survey show a negative correlation\footnote{Regions of active star formation, associated with high column densities, may show larger line widths
than quiescent regions due to feedback processes such as bipolar outflows and \textsc{HII} regions. 
Observations of specific molecular clouds from the literature may not be representative of all molecular clouds because
observations of active regions are over-represented compared to quiescent regions
due to selection bias in favor of astrophysically interesting observational targets.}, 
but most do as can be seen in figure \ref{coadded_profiles}\!\! that shows 
the non-dimensional, sample-average radial profile of the line width from 3683 clouds (\S\ref{Co-addition}).  

From the minimum value at the beginning of the profile ($R=0$) to the maximum value at the end ($R =20$ HWHM),
the line width increases by 40\%. About half the increase occurs within the cloud.

The corresponding sample-average radial profile of the $^{13}$CO integrated intensity, 
figure \ref{coadded_profiles}\!\!, also shows a continuous relationship between the
cloud and its surroundings. 
Two column densities from the Lane-Emden equation for hydrostatic equilibrium are included for comparison. The dotted line shows
the profile with a constant non-dimensional temperature. 
The long-dashed line shows the profile with the non-dimensional temperature proportional to
the square of the sample-average radial profile line width (figure \ref{coadded_profiles}\! {\it left}), 
proportional to the turbulent energy. 
There is no
statistical reason why the individual profiles should average to the solution of the Lane-Emden equation so precisely
as in figure \hbox{\ref{coadded_profiles}\!.}

The profile of the squared line width and the solution of the non-isothermal Lane-Emden equation can be
used to fit a poytropic equation of state $P(r) = K \rho(r)^\gamma$ with $\gamma=0.92$.

The interpretation of average hydrostatic equilibrium in the molecular ISM requires consideration. 
The Nyquist sampling of the GRS ensures that the data include
the full range of accessible spatial scales from the limit of the angular resolution at 46 arc seconds to the maximum extent
of the radial profiles at 48 arc minutes. The two average radial profiles are then averages across all spatial scales.  
On average, the turbulence on any scale is in hydrostatic equilibrium with 
the scales above and below. On average, the turbulent molecular ISM is neither collapsing nor expanding on any scale.
This is consistent with the low efficiency of star formation. At any time, only a small fraction, 0.6\%, of the molecular
ISM is gravitationally collapsing to form stars \citep{Evans_2021}. 

Because of the relationship between the turbulent dynamical time scale and the spatial scale, we can think of the averaging across
spatial scales as an averaging across dynamical time scales. Therefore, the property of average hydrostatic equilibrium
is a stationary or time-independent property of
the turbulence.  Because this property applies on all scales including the length scale of each molecular cloud, there is a
stationary component of hydrostatic equilibrium in each cloud as well
(figure \ref{individual_profiles}).

 The molecular ISM between clouds should also be in approximate hydrostatic equilibrium. As the distance
 increases from the concentration of mass in a molecular cloud, the hydrostatic equilibrium becomes
 dominated by pressure balance.

\begin{table*}[!ht]  
\caption{Log normal distributions of radius, column density, and line width}
\begin{tabular}{llcc} 
\hline
& radius, $R$ & col. density, $\Sigma$ & line width, $\sigma$ \\
& (pc) & (g cm$^{-2}$) &  (km s$^{-1}$)\\
\hline
      log                    &  $0.56 \pm 0.33$ 	& $ -1.97 \pm 0.27$	& $0.27 \pm 0.22$ \\
      linear                & $4.85 \pm 4.29$ 		& $0.013 \pm 0.001$ & $2.11 \pm 1.1$ \\
\hline
\end{tabular}
\label{table_lognormals}
 \end{table*}

\begin{table*}[!ht]
    \caption{Comparison of Cloud Properties, Median Values}
    \label{table1}
    \begin{tabular}{lcccc}
    \hline
     & {radius} & {col. density} & {$\Delta v$} & $\langle \Delta v \rangle^1$\\
      & (pc)  & (g cm$^{-2}$)  & (km s $^{-1}$) & (km s $^{-1}$)\\
      \hline
      This study                       & 3.8   &   0.0086   &    1.4  &   2.0 \\
      Rathborne et al. 2009     & 2.8   &  0.0083   &     1.2\\
      \hline
    \end{tabular} 
    \begin{tablenotes}
    \item[$^1$]{$\langle \Delta v \rangle $ indicates the mean of the velocity dispersions of all the pixels
    within the cloud radius, while $\Delta v$ in this study derives from the single pixel of peak integrated intensity.
    In Rathborne et al (2009), $\Delta v$  refers to the spectral FWHM defined by their
    segmentation algorithm and converted here to
    Gaussian width for comparison.}
 \end{tablenotes}
 \label{table_medians}
\end{table*}

\subsection{Virial Equilibrium in Pressure}\label{PVE}

\begin{figure}[!ht]
\includegraphics[width=3.5in,trim={0.85in 1.0in 9.0in 4.8in},clip]{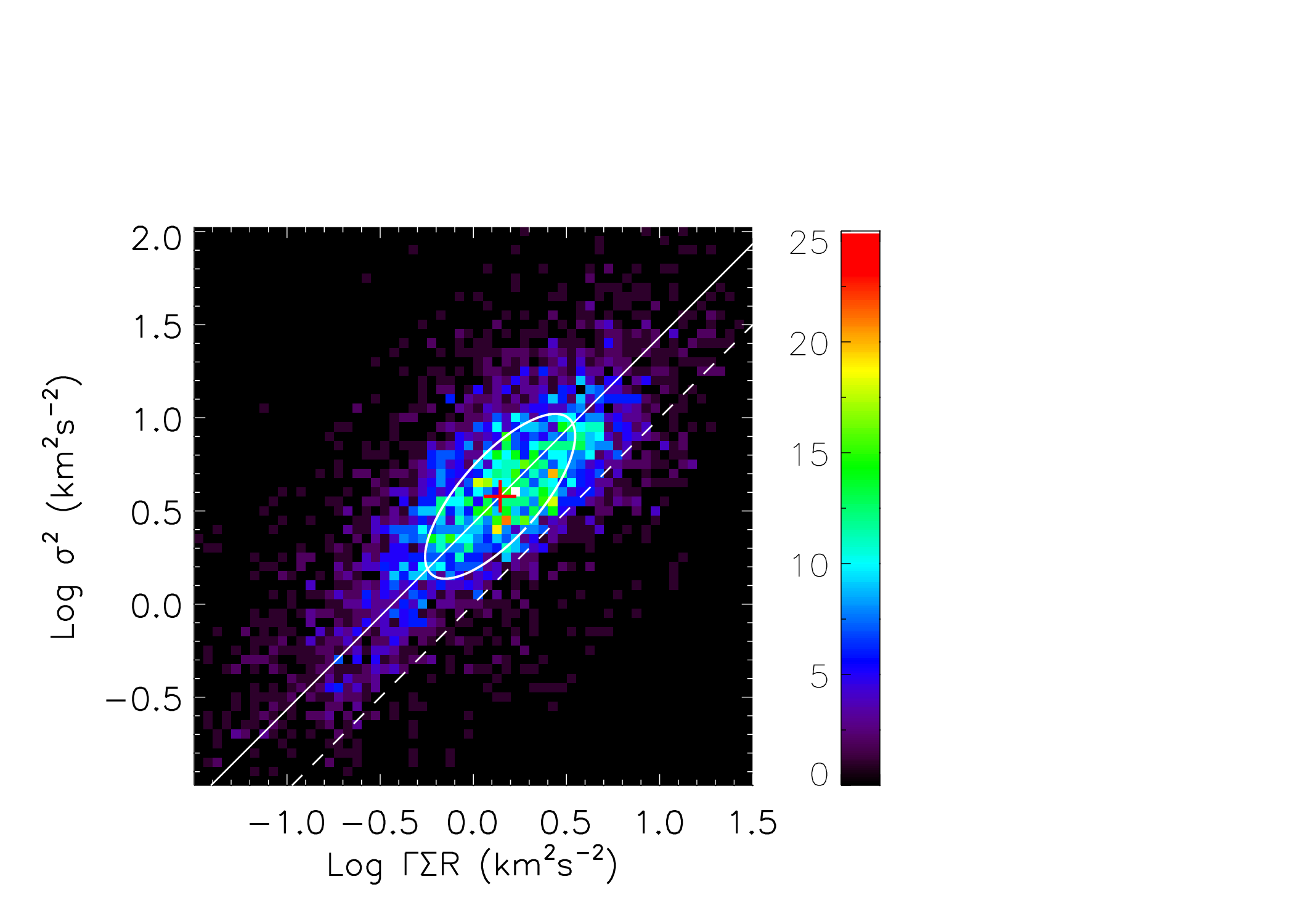}
\caption{
A 2-D histogram of the GRS clouds binned according to their gravitational potential (GE) and kinetic (KE) energies 
per unit mass with
colors representing the number of clouds per bin of size (0.05 $\rm km^2 s^{-2})^2$.  
The ellipse shows the standard deviations of the distribution  from a 2-D Gaussian fit (table \ref{table_pve_energy}\!\!)
with the mean marked with a cross. Constant pressure appears as a straight line in this space.
The solid line identifies the average external pressure consistent with the three term
virial theorem (equation \ref{ve1}) and the mean GE and KE.   The line of zero pressure is shown dashed.
}
\label{pve}
\end{figure}

 Figure \ref{pve}\!\! shows a 2-D histogram of 4190 GRS clouds (\S\ref{Cloud_ID})
 in the plane of the kinetic energy (KE) vs. the gravitational potential energy, (GE), both per unit mass,
 and the number of clouds in bins of (0.05 $\rm km^2\ s^{-2}) ^2$ in color. 
 The parameters of a 2-D Gaussian fit to the distribution are given in table \hbox{\ref{table_pve_energy}\!.} 
 An ellipse is shown with position angle $49^\circ$ and
 with semiaxes respectively equal to the variances of the energies, 0.06 and 0.02 $\rm km^2\ s^{-2}$.
 In the 2-D space of energy, the clouds are clustered in this elliptical distribution whose major axis is
 approximately aligned with the $45^\circ$ straight line of virial equilbrium with
 a constant pressure energy per unit mass (equation \ref{ve1}),
 $P_e = 1.17 \pm 0.16 \rm\ km^2\ s^{-2}$. 

\begin{figure}[!ht]
\includegraphics[width=3.5in,trim={0.2in 0.4in 2.3in 0.4in},clip]{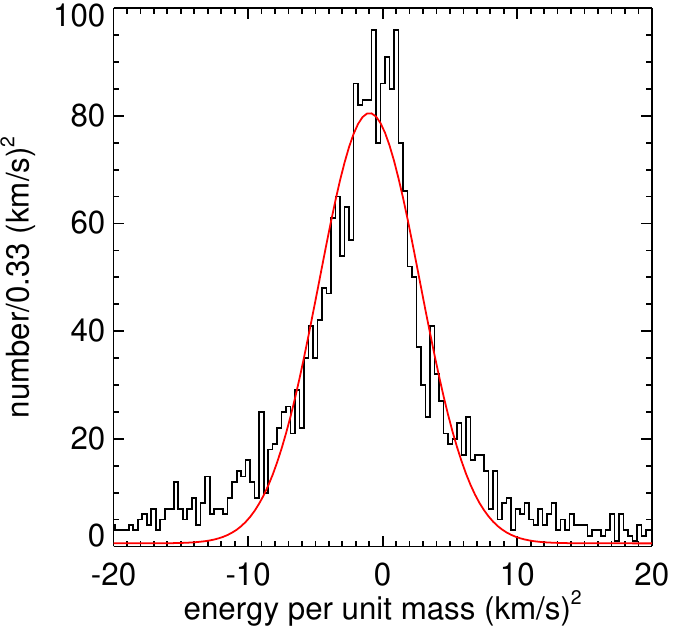}

\caption{
A 1-D histogram of the variable $\delta$ expressing the departure from time-dependent virial equilibrium from equation \ref{ve1}.
The departures have a ragged normal distribution indicating that the measured mean $\delta = -0.97 \pm 3.76 \ (\rm km^2\ s^{-2})$
is consistent with time-dependent virial equilibrium.
}
\label{v_histo}
\end{figure}

The local, external pressure energy 
can also be measured directly from the observed azimuthally-averaged line width 
at the cloud boundary, $\sigma^2(R)$
(equation \ref{ve1}).  The measure of virial disequilibrium (equation \ref{ve1}), 
$\delta = -0.97 \pm 13.76$ $\rm km^2\ s^{-2}$. 
The distribution of $\delta$ shown in figure \ref{v_histo} indicates that the difference from zero is not significant.
Thus the energies of the individual clouds are in  time-dependent virial equilibrium with their
local external pressure energies estimated directly from the $^{13}{\rm CO}$ line widths around the clouds.

The elliptical distribution of the KE
and GE is a property of the fluctuating component of the turbulence. 
The alignment of the major axis indicates a preference for combinations
of KE and GE that satisfy virial equilibrium with external pressures
given by equation \ref{ve1} and figure \ref{v_histo}\!\!.
A standard
population-lifetime argument then suggests that
the clouds are able to evolve toward virial equilibrium faster than the larger scale 
fluctuations in the turbulence perturb them out of equilibrium.

 \begin{table*}[ht!]  
    \caption{Parameters of the 2-D Gaussian distribution of clouds, energy per unit mass}
     \begin{tabular}{lcccc} 
\hline
		&	GE						& KE		& GE						& KE			\\
		&	$\rm (km^2 s^{-2})$ 		&$\rm (km^2 s^{-2}) $	& $(\rm ergs\ cm^{-3})$ 	&$(\rm ergs\ cm^{-3}$) \\
\hline
log		&$0.14 \pm 0.11$		&$0.58 \pm 0.28$	& $-11.26 \pm 0.57$		&$-10.82 \pm 0.57$\\	
linear	&$1.44 \pm 0.36$		&$4.66 \pm 3.33$	& $1.31 \pm 2.85\times 10^{-11}$		&$3.53 \pm 7.69\times 10^{-11}$\\
\hline	
\end{tabular}
\begin{tablenotes}
\item{Quantities refer to the unrotated coordinates parallel to the axes of figure \ref{pve}, GE and KE.}
\end{tablenotes}
 \label{table_pve_energy}
 \end{table*}

\subsection{External Pressure}

The comparable variability of the  PE outside the clouds and the KE and GE inside
the clouds (\S\ref{PVE}), suggests that the external PE is due turbulent fluctuations in the molecular ISM
on scales larger than the clouds.
In figure \ref{individual_profiles}\!\!, the perturbations from hydrostatic equilibrium 
at and beyond the cloud radius (2 HWHM) contribute to the variability of the confining external pressure energy. 

In units of energy per volume, the elliptical distribution has semiaxes of 
0.78 and 0.21 $\rm ergs\ cm^{-3}$, and
the average external pressure,
 $ P_e  = 2.12 \pm 1.96 \times 10^4\ k_B\ {\rm K\ cm}^{-3}$. 
This pressure is comparable to estimates of the  
pressure of the ISM at the mid-plane of the Galaxy, $\sim 1\times 10^4\ k_B\ \rm K\ cm^{-3}$ 
\citep{Bloemen_1987,Elmegreen_1989,Boulares_1990}. 
The average external pressure represented by the solid line in figure \ref{pve} may be a stationary property of the turbulent
molecular ISM that
is imposed as an outer boundary condition by the pressure of the multi-phase ISM.

\subsection{Larson's Laws}\label{Larson}

Based on data from the early days of radio frequency spectral line observations, \citet{Larson_1981} 
proposed that the line width and
number density averaged within molecular clouds scale with the cloud length, $L$, 
as $\sigma \propto L^{0.38}$ and $n \propto L^{-1.10}$, respectively,
over 3 orders of magnitude in length. The two relationships straddle the dichotomy between turbulence and stability
in the ISM. Turbulence is indicated by the exponent of the first scaling relationship, between 1/3  consistent with incompressible turbulence
and 1/2 consistent with compressible turbulence \citep{Padoan_1995,Kritsuk_2013}. Stability
is indicated by second scaling relationship because virial equilibrium would be implied if the exponent of the first scaling relationship were actually 1/2 
and the column density "nL $\sim$ constant" \citep{Larson_1981}. The validity of these scaling relations has been questioned on various grounds
(e.g. \citet{Scalo_1990,VS_1997,BP_2019}). 

Few concepts in the study of the molecular ISM have generated as much
confusion as the interpretation that the near zero regression coefficient in the column density vs. size 
relationship implies that "$nL \sim$ constant" meaning that clouds of all sizes have the same average column density.
The many follow-up studies, as evidenced by 2490 citations to \citet{Larson_1981}, that 
include those seeking to verify, disprove, or improve, have not cleared up the confusion nor offered a satisfactory explanation of 
the origin and implications of the two scalings. We return to this question after discussing our results of
regressions of the properties of the GRS clouds.

\subsubsection{Correlations}

\begin{figure*}[!ht]
\begin{tabular}{cc}
\includegraphics[width=3.4in,trim={0.2in 0.4in 2.3in 0.4in},clip] {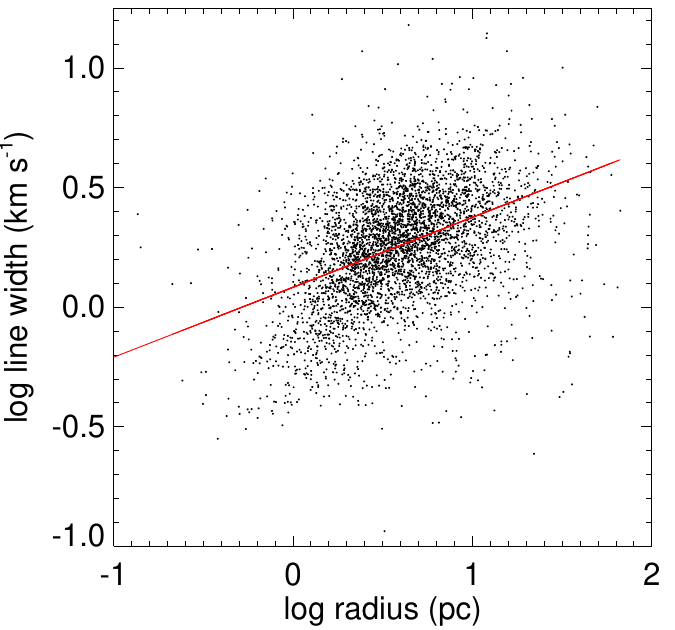} &
\includegraphics[width=3.4in,trim={0.2in 0.4in 2.3in 0.4in},clip] {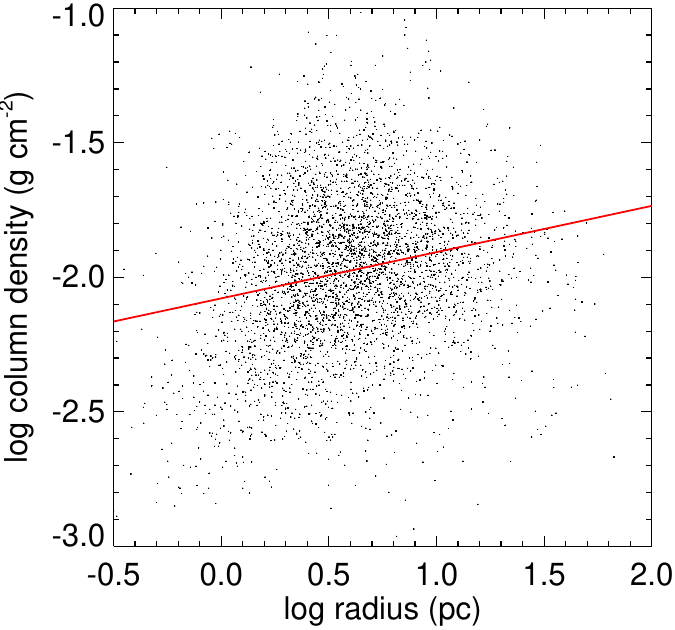}\\

\end{tabular}
\caption{
        {\it Left:} Line width vs radius and {\it right:} column density vs radius both with regressions (solid red line). 
}
\label{regression1}
\end{figure*}

Figure \ref{regression1}\!
shows the regressions of $\sigma$ vs $R$ 
and $\Sigma$ vs $R$  with the results listed in \mbox{table \ref{regression_results}\!\!}.
Because of the large sample size (4190 clouds), standard tests on the significance, p-value, F-test, or maximum likelihood,
all find a vanishing small probability that these fits could arise by chance. Therefore, we evaluate the 
statistical significance by a different method. 

The root mean square errors (RMSE)
in both cases \mbox{(table \ref{regression_results}\!)} are similar to the log standard deviations of the dependent variables (table \ref{table_lognormals}\!\!).  
With $Y$ as either of the dependent variables, $\sigma$ or $\Sigma$, and $X=R$, 
the RMSE from the regression, $Y = b_0 + b_1X_1$,
is seen from the two tables to be essentially the same as the standard deviation of $Y$.  Therefore, the 
expected value for the $i$th member of the sample, $\hat Y(i) =b_0 + b_1 X_1(i) \pm {\rm RMSE}$, is  
no better estimate than the expected value of $Y(i)$ without the
regression which is the mean of the sample $\mu_Y \pm \sigma_Y$ where $\sigma_Y$
is the standard deviation of $Y$. 

\begin{figure*}[!ht]
\begin{tabular}{cc}
\includegraphics[width=3.4in,trim={0.2in 0.4in 2.3in 0.4in},clip] {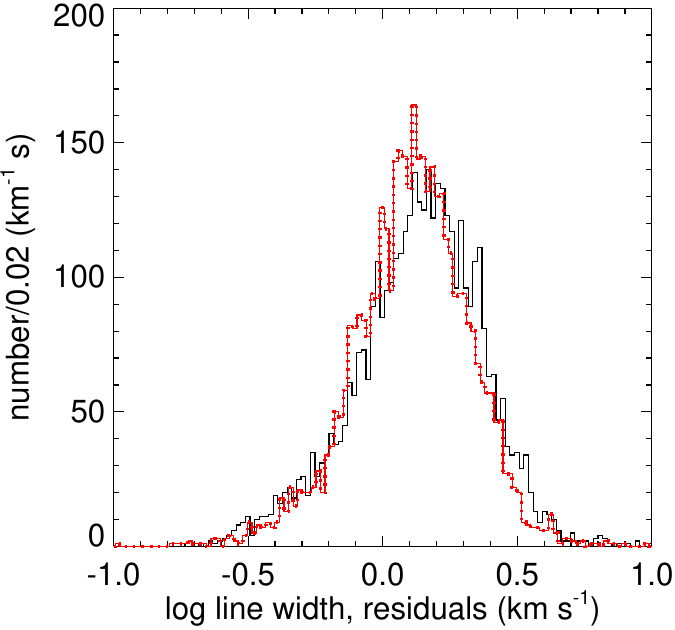} &
\includegraphics[width=3.4in,trim={0.2in 0.4in 2.3in 0.4in},clip] {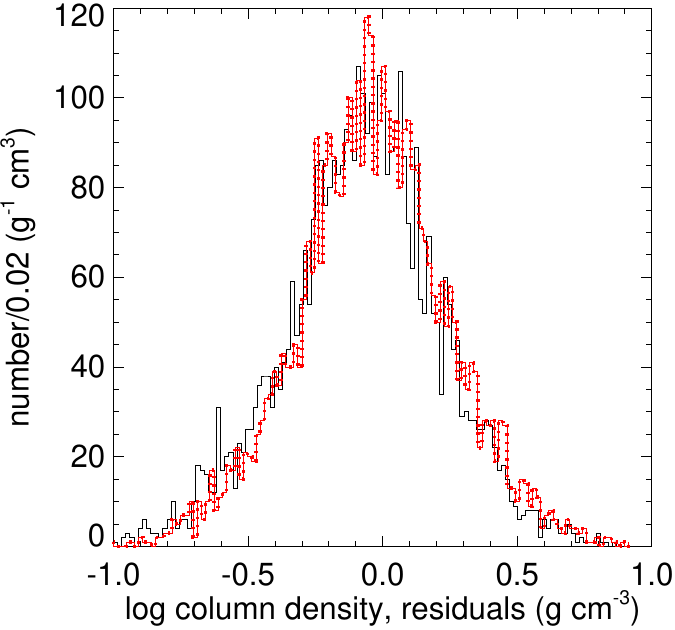} \\

\end{tabular}
\caption{
        Histograms of log line width {\it Left}  and log column density {\it right} 
        both with histograms of the regression residuals. The cloud properties are shown in black and the residuals in red
        with horizontal hash marks. The histograms of cloud properties are the same as in figure \ref{histograms}.
        The residuals derive from the regressions in (figure \ref{regression1}).
}
\label{residuals1}
\end{figure*}

The point is demonstrated graphically in figure \ref{residuals1}\! which
compares the distributions of the dependent variables, $\sigma$
and $\Sigma$, with the distributions of their respective regression residuals. In both cases the 
distributions of the variable and the residuals are essentially the same. If a regression were
significant, the mean of the residuals would shift left to a more negative number in the log space, and the width
of the distribution (RMSE of the regression) would be narrower. Figure \ref{residuals1}\!\! ({\it left}, $\sigma$ vs. R)
shows a slight shift of the two distributions consistent with ${\rm RMSE} < \sigma_\sigma$ although both equal 0.22 km s$^{-1}$.
Because of the large sample size, the
Kolmogoroff-Smirnov test for the equivalence of two distributions is inconclusive.

\begin{table*}[!ht]  
    \caption{Regressions}
     \begin{tabular}{llccc} 
\hline
					&Regression coeff. {b} 	& Std. err. {$\sigma_b$} & {correlation coeff.}  &RMSE \\
\hline
$\sigma$ vs. $R$				& 0.29 (km s$^{-1}$ pc$^{-1}$)		& 0.01 (km s$^{-1}$)		& 0.71		&0.22	(km s$^{-1}$)	\\
$\Sigma$ vs. $R$			& 0.17 (g cm$^{-2}$ pc$^{-1}$)		& 0.01 (g cm$^{-2}$)		& 0.21		&0.28	(g cm$^{-2}$) \\
\hline
\end{tabular}
\label{regression_results}
 \end{table*}

\begin{figure}[!ht]
\begin{tabular} {c}
\includegraphics[width=2.52in,trim={0.0in 0.4in 2.4in 0.4in},clip]{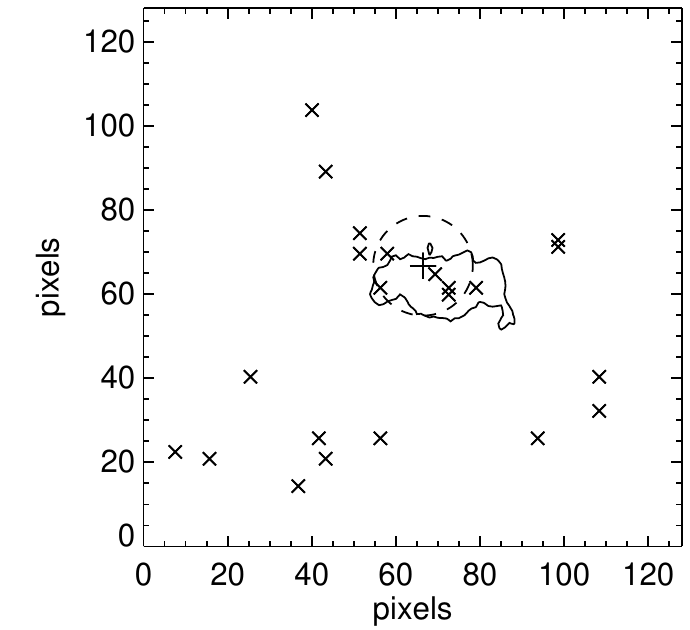} \\
\includegraphics[width=2.52in,trim={0.0in 0.4in 2.4in 0.4in},clip]{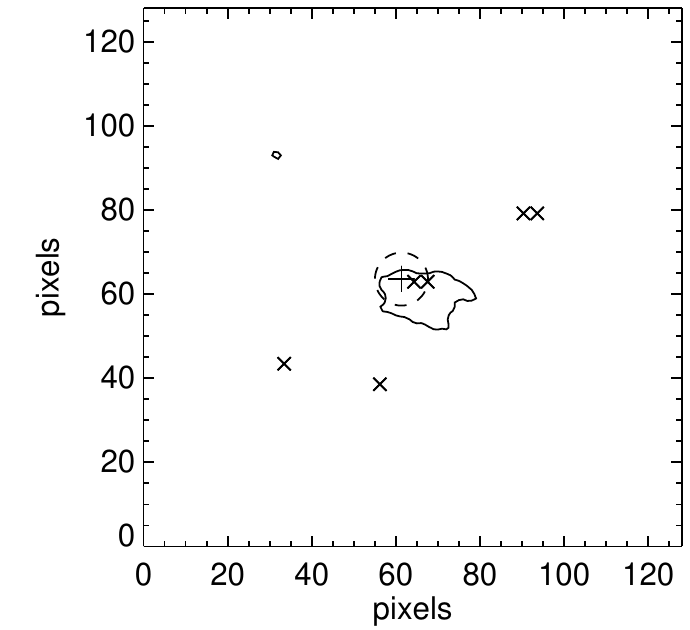} \\
\includegraphics[width=2.52in,trim={0.0in 0.4in 2.4in 0.4in},clip]{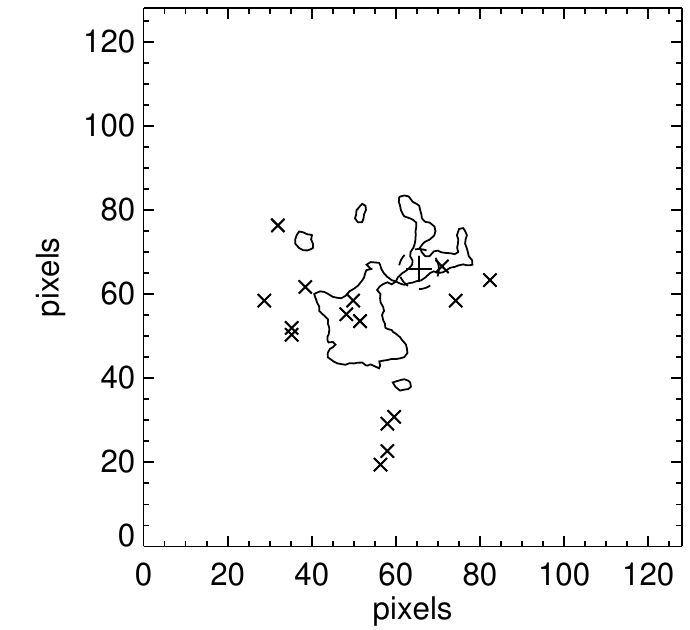} \\
\end{tabular}
\caption{
Contour maps of the $^{13}$CO integrated intensity showing the complexity of cloud shapes. 
The single contour level is the half-power level
of the cloud marked  "+".
The dashed circle shows the HWHM radius. Nearby clouds with similar VLSR are marked
as "x".  
The examples {\it top} to {\it bottom} are clumps 5, 45, and 34 in \citet{Rathborne_2009} with complexities
of  1.5, 1.6, and 3.9, respectively.}
\label{complexity}
\end{figure}

\mycomment{
\begin{figure*}[hbt!]
\begin{tabular} {lll}
\includegraphics[width=2.5in]{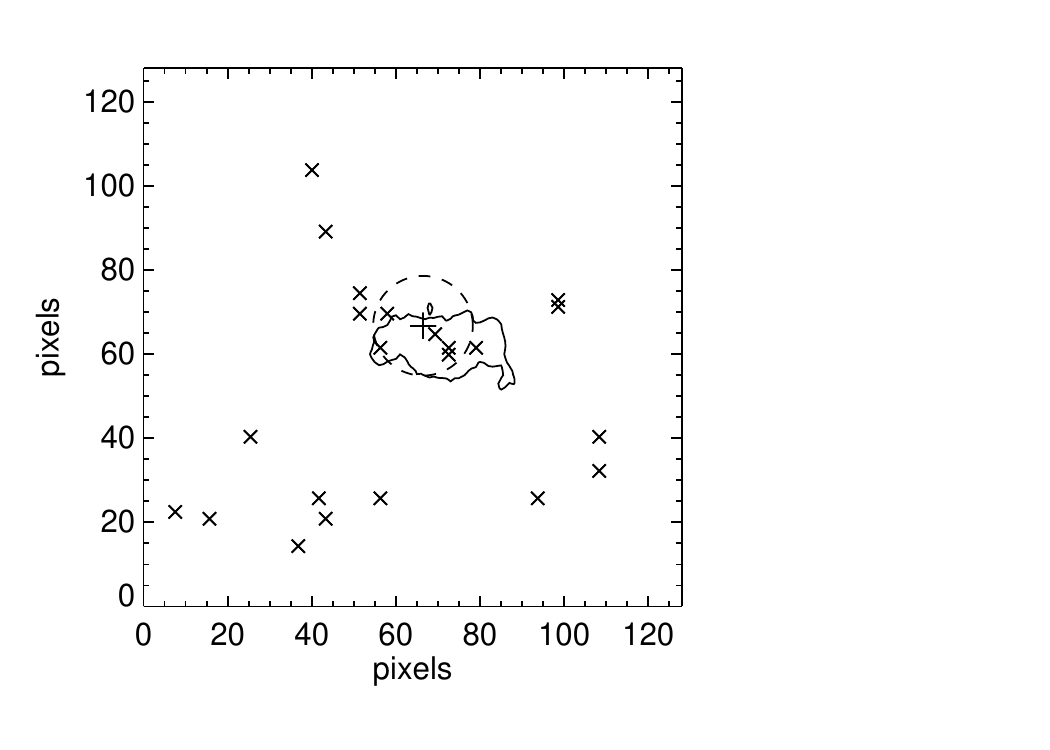} &
\includegraphics[width=2.5in]{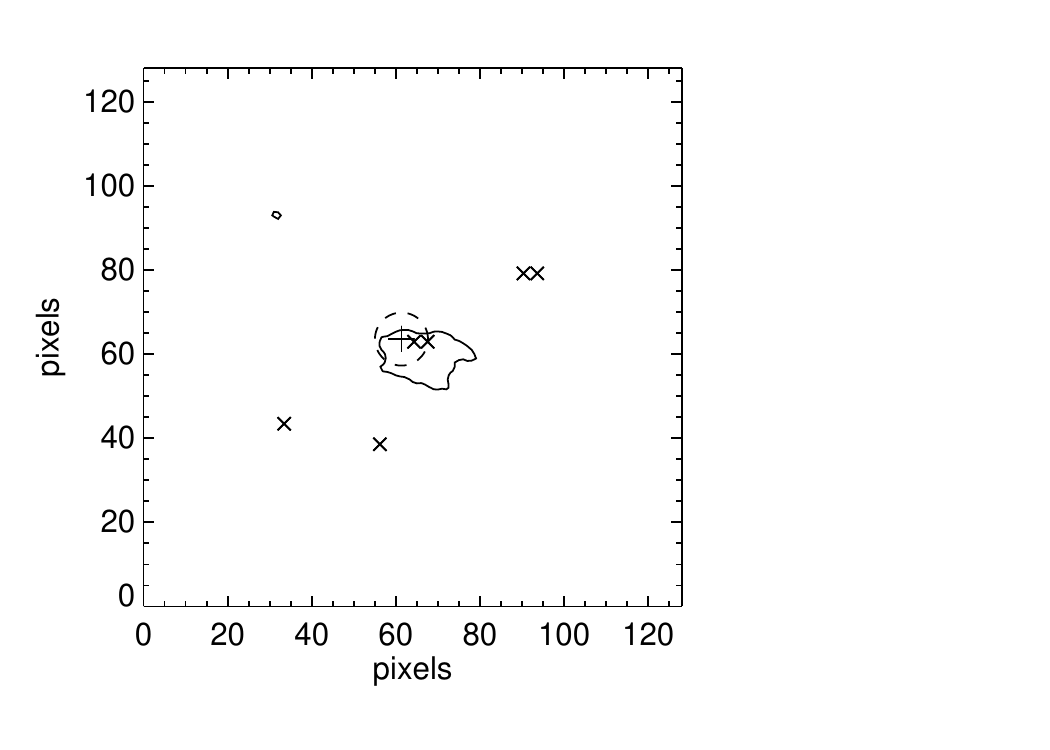} &
\includegraphics[width=2.5in]{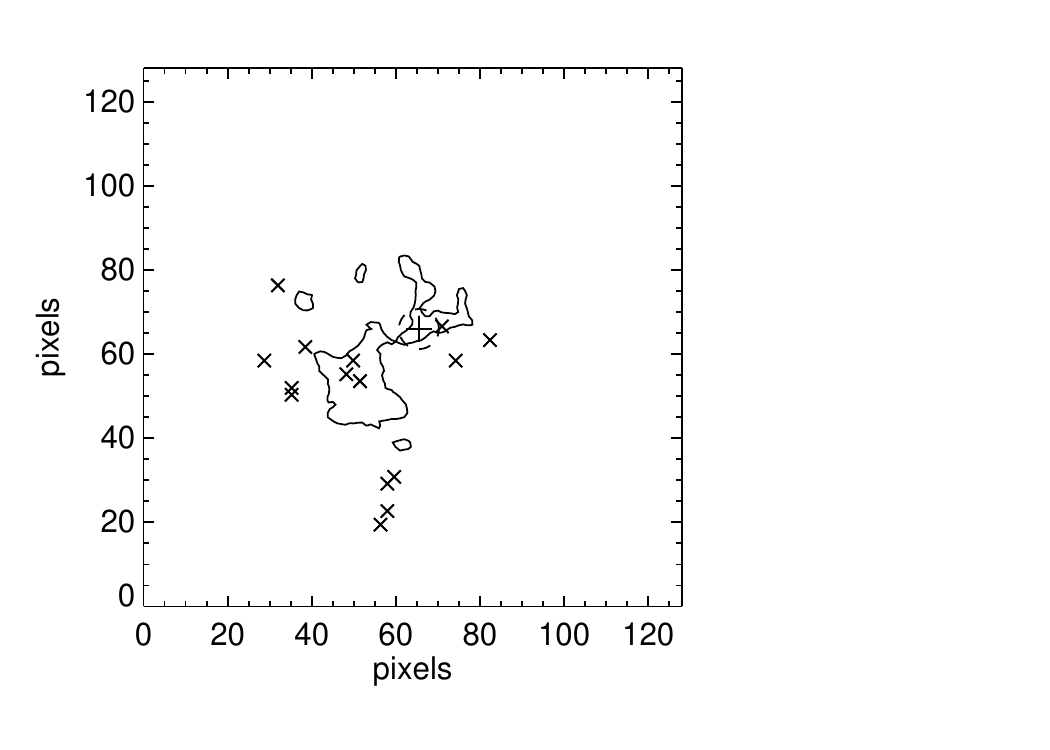} \\
\end{tabular}
\caption{
Contour maps of the $^{13}$CO integrated intensity showing the complexity of cloud shapes. 
The single contour level is the half-power level
of the cloud marked  "+".
The dashed circle shows the HWHM radius. Nearby clouds with similar VLSR are marked
as "x".  
The examples {\it top} to {\it bottom} are clumps 5, 45, and 34 in \citet{Rathborne_2009} with complexities
of  1.5, 1.6, and 3.9, respectively.}
\label{complexity}
\end{figure*}
}

\subsubsection{Uncorrelated or constant column density}

We return to the interpretation of a near zero value for the coefficient $b_1$ in the regression of $\Sigma$ and $R$.
If the dependent variable $\Sigma$ were strictly constant aside from measurement errors, then $\Sigma$ and $R$ would be trivially uncorrelated. 
The interpretation of whether $\Sigma$ is best described as constant
or uncorrelated depends on the relative variances of $\Sigma$ and $R$ in the context of the 
physics under consideration, 
for example, virial equilibrium. Omitting the pressure term, equation (\ref{ve1}) for virial equilibrium in terms
of the energy per unit mass written 
in non-dimensional units is,
\begin{equation}
s^2 - rS = 0
\end{equation}
with these substitutions, $s= \sigma / \sigma_0$, $r = R/R_0$, and $S = \Gamma \Sigma R_0/\sigma_0^2$
for any $\sigma_0$ and $R_0$, for example,  $\sigma_0 = \mu_\sigma$ and $R_0=\mu_R$.
We find the non-dimensional standard deviations of $s,\ r,\ S = 0.88,\ 0.54,\ 0.15$. Since the standard deviations of the
non-dimensional column density, $S$, and radius, $r$, are of the same order, the relationship between $\Sigma$ and $R$
is better described as uncorrelated. Constant might have been appropriate if the two were of different magnitude such that the standard
deviation of the non-dimensional GE, $rS$, were dominated by standard deviation of $r$.
The observational evidence from the GRS clouds contradicts the interpretation that $b_1\sim 0$ implies 
constant, average column density for clouds of all sizes.

\subsubsection{Autocorrelation}

The correlation coefficient in regressions with one independent variable has a simple interpretation with zero 
indicating no correlation and and  a value of one indicating perfect correlation.
The regression coefficient, $b_1$, relating column density, $Y$, and radius, $X_1$, can also be 
derived in a regression where $Y = n$, the number density \citep{Larson_1981}, or $Y = M$ the mass 
\citep{Lombardi_2010}, as $b_{1,\Sigma} = b_{1,n} +1$ or $b_{1,\Sigma} = b_{1,M}-2$, respectively. 
However, in these regressions, the correlation
coefficient always has a higher value closer to one because of the autocorrelation of $n$ or $M$ with radius.

Observationally, the mass is derived from sum of the $^{13}$CO integrated intensities
of all pixels within the cloud. Since each
pixel represents an area, the autocorrelation of mass and radius is inherent. 

The number density of molecular clouds
cannot be derived in a simple way from observations of spectral line intensity because of
the dependence of the intensity on the unknown path length through the cloud along the line of sight. 
The number densities in figure 5 of Larson (1981)
that show a correlation with cloud size are derived from either the mass or column density reported by
the observational studies cited and then by conversion by the appropriate power of $R$. 
The apparent correlation in figure 5 of \citet{Larson_1981} is 
necessarily affected by autocorrelation.

\subsubsection{Other correlations}

We find no interesting scaling relations in  multiple regressions over a grid of variables other than the
correlation between KE and GE (\S\ref{PVE}). 
Correlations that turned up were either due to
autocorrelation or could be related back to the correlation of the energies. For example, in the scaling relationship suggested in
Heyer et al. (2009), the variables $\sigma ^2/R$ and   $\Sigma$  are the KE and GE per unit volume
after cancelling a common factor of $\Sigma$. 

\subsection{Complexity}\label{Complexity}

Most of the GRS clouds have complex, non-circular shapes inconsistent with the 2-D projection of the
3-D spherical, minimum-energy configuration of hydrostatic equilibrium. Their shapes appear to be significantly
affected by local turbulent fluctuations. We quantify the non-circular asymmetry
with a complexity parameter defined as the ratio of the path length of the half-power contour around the peak integrated intensity
to the circumference of a circle at the half-power radius.
Figure \ref{complexity}\! shows three example clouds with complexities of 1.5, 1.6 and 3.5. 
Although an elliptical
cloud with a 2:1 axial ratio would have a complexity of 1.6, the examples in figure \hbox{\ref{complexity}\!,} ({\it left}) and ({\it mddle}),
show that 
cloud shapes with this complexity may already
be more complex than elliptical. 
Also clouds with complex structure generally contain more than one cloud (or clump)  shown as X's. 
The distribution of complexity is log normal, and 89\% of the clouds have a complexity $> 1.2$. The complexity is not
correlated with any other cloud properties such as the ratio KE/GE.

\subsection{Column Density Probability Distribution Functions}\label{PDF}

Previous studies of low-mass star forming regions \citep{Kainulainen_2009} 
as well as some giant molecular clouds \citep{Schneider_2015} suggest that the 
column densities measured by infrared absorption have
probability distribution functions (PDFs) that can be fit with a log normal at lower column densities and a
power-law  at higher column densities. The change in functional behavior
is attributed to 
contraction of the clouds and star formation at the highest densities.  This observational 
description is not so clear if the column densities are
determined from molecular lines, even those that are
typical tracers of high-density gas. In two studies \citep{Wang_2020, Schneider_2016} column density PDFs 
from HCO$^+$(1-0), HCN(1-0) HCN(1-0), and CS(2-1)
in giant molecular clouds
show power-law slopes at higher densities, those from $^{13}$CO(1-0) do not, while the N$_2$H$^+$(1-0) PDFs are inconsistent.  
Possible complications to the interpretation of PDFs based on molecular line observations include abundance variations due to
chemistry and saturation of the spectral lines due to optical depth.

\begin{figure}[!ht]
\includegraphics[width=3.0in,trim={0 0 2.3in 0},clip] {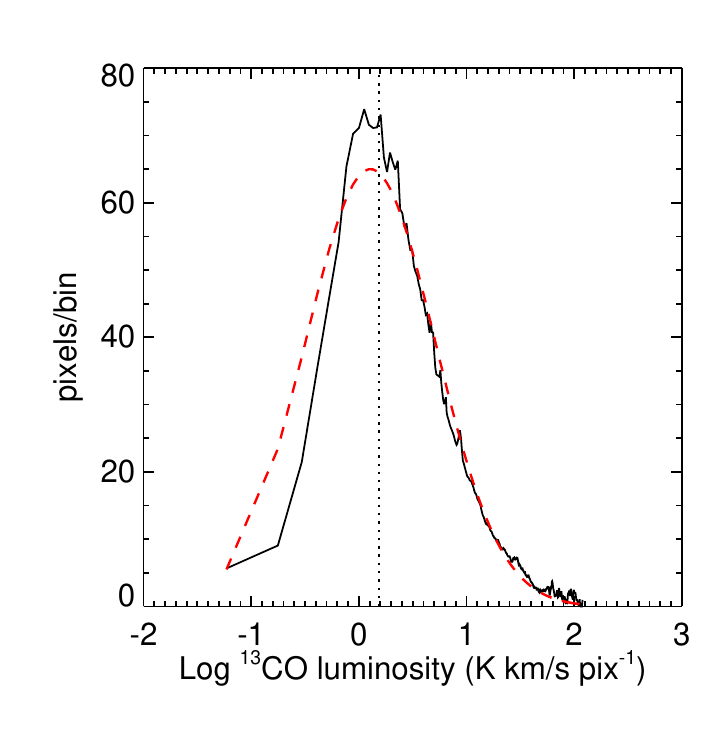} 
\caption{
{\it Left:} Sample average column density PDF of the clouds on a log scale.
The observational noise level is shown by the vertical, dotted line, and the
best fit log normal Gaussian by the long-dashed line.}
\label{fig_pdfs}
\end{figure}

Although we lack star-formation
tracers to identify contracting regions of the GRS clouds, 
we can compare column density PDFs from gas inside and outside of our 2 HWHM cloud boundaries.
The PDFs of individual clouds are too noisy to be useful, so we resort to rescaling and averaging.
Figure \ref{fig_pdfs}\! shows the sample-average PDF of pixels within
the clouds. 
The sample-average PDF outside the clouds is essentially
the same   and also
consistent with a log normal distribution. This result is consistent with the two $\rm ^{13}CO$ studies cited above.

However, a consideration of the completeness limit set by the  observational noise 
allows a second interpretation of the PDF.
We estimate the noise level in the sample average as $(2\langle\sigma_v\rangle/\Delta_v)^{1/2}\sigma_{T}$
where $\langle\sigma_v\rangle$ is the average velocity dispersion of the clouds, $\Delta_v$ is the spectral resolution, and
$\sigma_T$ is the observational noise in units of the antenna temperature. The estimated 
$3 \sigma$ noise level is marked on figure \ref{fig_pdfs}
as a vertical line. 
The PDF could also be interpreted with a single power law that is fit to the distribution to the right of the 
completeness limit if the
falloff in the number of clouds on the left side is due to incompleteness.

\section{Discussion}\label{Discussion}

The negative correlation of the $^{13}$CO integrated intensity and spectral line width (\S\ref{Hydrostatic}) indicates
a lower turbulent energy per unit mass at higher column densities. 
The explanation depends on which turbulent processes are responsible
for the formation of molecular clouds. 

If the clouds
are formed by compression between turbulent flows, 
then the clouds could have lower line widths than the surrounding turbulence because the
compressed regions are stagnation points with lower velocity dispersions \citep{Klessen_2005}.
 
The explanation is different if the clouds are formed by the unstable growth of perturbations
in energy due to a turbulent cooling instability \citep{Keto_2020}. 
The scale-dependent cooling rate, $\sigma/R$, results in lower turbulent velocities on the smaller scales within the clouds
and with respect to their surroundings.
 
However, there is a competing effect. Cooling, in combination with self-gravity and virialization,
results in a negative heat capacity whereby the turbulent energy is
rebalanced with the increased gravitational energy resulting from the contraction due to the loss of energy.
Dissipation and contraction thus work at cross purposes
with respect to the turbulent velocity dispersion.

The dissipation time scale and the 
dynamical time scale of the turbulence are both on the order of the crossing time, but different processes
are involved, and the two  need not be the same
within a factor of order unity. 
The observed decrease in the turbulent velocity dispersion with
spatial scale indicates that the turbulent dissipation time scale is shorter than the dynamical
time scale that mediates the virialization. 
This difference in the two time scales allows perturbations to grow before they are erased by virialization. 
The turbulent molecular ISM described in this study is
suitable for the turbulent cooling instability.

Clouds that contract to 
a critical density for dynamical stability with respect to a Jeans criterion either collapse to
star formation or fragment into smaller clouds. The dispersion relation for the growth rate as a function
of the dissipation rate indicates that fragmentation is preferred for faster rates of dissipation. 
The process is scale free and results in a fragmentation cascade as each newly formed
fragment follows the same evolutionary path.

The fragmentation
process cannot be followed by the analysis of a snapshot in time of the individual clouds, their
surroundings, and their average properties. Analyzed differently, the GRS data may be
helpful. A hierarchy of the GRS clouds is already suggested
in the analysis by \citet{Rathborne_2009}. That study defines cloud complexes (their clouds) 
as associations of smaller clouds (their clumps)
within a region of similar Galactic coordinates and VLSR ({\it l,b},VLSR). If this hierarchy represents two levels
in a fragmentation cascade, some information on the fragmentation process may be gained from the study of the relationship of
their clouds and clumps.

\section{Conclusions}\label{Conclusions}

We analyzed the $\rm ^{13}CO$ spectral line data of the GRS survey with new methods and provide new 
observational information on the properties of the turbulent molecular ISM that hold regardless of interpretation. However, in the absence of
interpretation, the results showing hydrostatic and virial equilibrium appear paradoxical 
within a turbulent molecular ISM that is characterized by instability. The paradox can be resolved by consideration of
the variation in the time scale for the evolution of turbulence at different spatial scales and consideration of
the spatial scales of clouds within the surrounding molecular ISM. 

Examination of Larson's scaling relations for column density and line width with radius finds neither
correlation is significant. The inference of constant column density with cloud size is a
misinterpretation of the lack of correlation.

\subsection{Observable properties of turbulence in the molecular ISM}\label{Discussion_Properties}

The analysis of the GRS data provides specificity on the conditions of the turbulent molecular ISM. 
We identify both stationary and fluctuating properties.
These observational properties  should be useful for assessing theoretical models of 
turbulence in the molecular ISM \\
1) There is an overall average hydrostatic equilibrium across all scales in the turbulence.
This is a stationary property that allows for local fluctuations and 
does not require any region to be in hydrostatic equilibrium. \\
2) Regions with higher column density have lower spectral line widths and lower turbulent kinetic energy per unit mass.  \\
3) The kinetic and gravitational energies within a cloud are virialized with a ratio set by the local pressure energy 
around each cloud. The three energies are fluctuating components of the turbulence.  \\
4) The ratio of the mean kinetic and gravitational energies per unit mass within clouds is $\langle 2KE\rangle / \langle GE \rangle \sim 2.9$. \\
5) The average excess kinetic energy interpreted as a pressure (energy per unit volume)
is comparable to the average 
pressure of the multiphase ISM at the Galactic mid-plane. This is a stationary property of the turbulence.
The mid-plane pressure of the Galaxy may be a boundary condition for the molecular phase of the turbulence. \\

\bibliography{GRS_2024_08_27_REV}

\section{Supporting Information}

Some results of this analysis are available on the Harvard Dataverse: Keto, Eric, 2024, "Replication Data for: Scales of Stability and Turbulence in the Molecular ISM", https://doi.org/10.7910/DVN/5Y0C4E, Harvard Dataverse, V1 
This additional data includes a catalog of the cloud properties such as radius, velocity dispersions, column densities, and mass. The additional
data also includes the azimuthally-averaged radial profiles of line width and column density as well as the PDFs of the pixels within and outside the
cloud boundaries.

 \end{document}